\documentclass[a4paper,12pt]{article}
\pdfoutput=1 

\usepackage{jheppub} 

\usepackage[T1]{fontenc} 

\usepackage{xcolor}
\usepackage{amsmath}
\usepackage{latexsym}
\usepackage{amssymb}
\usepackage{graphicx}
\usepackage{physics}
\usepackage{slashed}
\usepackage{float}
\usepackage{bm}
\usepackage{subcaption}

\title{\boldmath Remarks on Fermions in a Dipole Magnetic Field}

\author[a]{Jeff Murugan,}
\author[a]{Jonathan P. Shock,}
\author[a]{Ruach Pillay Slayen}

\affiliation[a]{The Laboratory for Quantum Gravity \& Strings,\\ Department of Mathematics and Applied Mathematics,\\
University of Cape Town, Private Bag, Rondebosch 7700, South Africa}

\emailAdd{jeff.murugan@uct.ac.za}
\emailAdd{jon.shock@gmail.com}
\emailAdd{slyrua001@myuct.ac.za}

\abstract{This work is a continuation of our recent study of non-relativistic charged particles, confined to a sphere enclosing a magnetic dipole at its center \cite{Murugan:2018hsd}. In this sequel, we extend our computations in two significant ways. The first is to a relativistic spin-$\frac{1}{2}$ fermion and the second concerns the interpretation of the physics. Whereas in \cite{Murugan:2018hsd} we speculated on the possibility of observing such condensed matter systems in the astrophysics of extreme magnetic sources such as neutron stars, the physical systems in this study are more down-to-earth objects such as a $\mathsf{C}_{60}$ fullerine enclosing a current loop. We unpack some of the details of our previous analysis for the spinless fermion on the dipole sphere and adapt it to solve the eigenvalue problem for the single-particle Dirac Hamiltonian. In the strong-field/small-radius limit, the spectrum of the spin-$\tfrac{1}{2}$ Hamiltonian, like the spinless case, exhibits a Landau level structure in the $|m|\ll Q$ regime. It features a new, additional (approximately) zero-energy lowest Landau level which persists into the $|m|<Q$ regime. As in the spinless system, the spectrum exhibits level-crossing as the strength of the magnetic field increases, with the wavefunctions localising at the poles in the strong-field/small-radius limit.}

\begin{document} 
\maketitle
\flushbottom

\section{Introduction}
\label{sec:intro}

The study of electronic matter in transverse magnetic fields has proven to be a fertile ground for the investigation of the quantum phases of matter. Directly, or otherwise, it has led to the discovery of a number of remarkable phenomena, from the Quantum Hall Effect (and its many variants) to topological quantum matter. While the degrees of freedom living on the boundary of such systems - the edge-states - play a key role in the physical properties of these examples, it is sometimes necessary to isolate the bulk degrees of freedom. One efficient device to do this was introduced by Haldane in \cite{Haldane:1983xm} (see also \cite{Jain} for a pedagogical elaboration) by putting the 2-dimensional system on the surface of a round 2-sphere and realising the perpendicular magnetic field by a magnetic monopole located at the origin of the $S^{2}$. The resulting system is known as the {\it Haldane sphere}. Another feature of 2-dimensional electronic matter in a constant (and perpendicular) magnetic field is the enormous degeneracy in energy levels, a key aspect of the Fractional Quantum Hall Effect (FQHE). These Landau levels persist in going from the planar problem to the Haldane sphere, but the degeneracy is lifted as soon as the magnetic field is not spatially homogeneous \cite{Grosse:1993np}. In the planar case, this result is readily understood, at least in the case of an axially symmetric magnetic field monotonically changing away from the origin. Eigenstates of the magnetic Schr\"odinger operator $H = (-i\bm{\nabla} - \bm{A})^{2}$ are labelled by angular momentum quantum number $m$ and Landau level number $n$, with states with larger $m$ values located further away from the origin. Consequently, eigenstates encountering a monotonically decreasing (increasing) magnetic field will exhibit a decreasing (increasing) set of eigenvalues $E_{m,n}$ as a function of $m$   within a given Landau level. For the lowest ($n=0$) Landau level, this can be proven exactly \cite{Grosse:1993np} while for $n>0$, one must be satisfied with a perturbative treatment. \\

\noindent
Producing such inhomogeneous magnetic fields in a laboratory setting is not easy. One possibility is to use the magnetic fields trapped in the core of vortices in type II superconductors which exhibit such a monotonically decreasing field strength away from the vortex core. Since a typical 200-300 \AA\, core-width vortex traps a 1-2 T magnetic field, reasonably large magnetic fields can be produced. However, boundary conditions are difficult to treat in such setups \cite{Bander:1989iu}. An alternative is to confine the electron gas to a curved surface, a sphere of unit radius say, in a constant ambient magnetic field $\bm{B}_{c} = B_{0}\bm{\hat{z}}$ whose normal component at any point $(\theta,\phi)$ on the sphere is given by $B_{0}\cos\theta$ and monotonically decreases from the pole to the equator. \\

\noindent
In a recent paper \cite{Murugan:2018hsd}, we introduced another device that combines these approaches by replacing the single monopole of the Haldane sphere with {\it two} monopoles with charges $+b$ and $-b$, separated by a distance $l$ and aligned along the $z$-direction. In the limit $l\to 0$ with $bl$ fixed, the resulting magnetic field,
\begin{eqnarray}\label{Bfield}
   \bm{B} = \frac{|\bm{\mu}|}{r^{3}}\left(2\cos\theta\,\widehat{\bm{r}} 
   + \sin\theta\, \widehat{\bm{\theta}}\right)\,,
\end{eqnarray}
is identical to that produced by a current loop enclosing a flat region with area $A_{\rm loop}$ in the $xy$-plane. In either case the magnetic moment is aligned in the positive $z$-direction with $|\bm{\mu}| = IA_{\rm loop} = bl$ and associated vector potential
\begin{eqnarray}\label{pot}
   \bm{A} = \frac{1}{r^{2}}\bm{\mu}\times\widehat{\bm{r}} = \frac{|\bm{\mu}|}{r^{2}}\sin\theta\,\widehat{\bm{\phi}}\,.
\end{eqnarray}
For comparison, the vector field associated to the constant magnetic field $\bm{B}_{c}$ above is, 
\begin{equation}\label{potc}
    \bm{A}_c = \frac{B_0}{2}r\sin\theta\widehat{\bm{\phi}}.
\end{equation}
Evidently, for particles confined to the surface of a sphere of radius $r=R$, these potentials (\ref{pot}) and (\ref{potc}) are proportional, and will yield qualitatively similar results, a point that we will return to at various points in this article. This {\it dipole sphere} problem is  relevant for quantum matter in astrophysical settings such as the ultra-dense atmospheres encountered in neutron stars, offering the tantalising possibility of probing condensed matter in sustained magnetic fields orders of magnitude stronger than anything produced terrestrially. Closer to home, it is also clearly of interest in the study of other compact configurations such as a $\mathsf{C}_{60}$ fullerene  enclosing a magnetic dipole or one of its larger allotropic cousins like $\mathsf{C}_{540}$, fullerite. Finally, and more speculatively, it is anticipated that the spectral problem for the Schr\"odinger (or Dirac) operator for a charged particle confined to a sphere punctured by a dipole magnetic field has some bearing on the physics of massless charged fermions moving in a magnetically charged black hole background \cite{Maldacena:2020skw},
\begin{eqnarray}\label{magBH}
   ds^{2} &=& -\left(1 - \frac{2MG_{N}}{r} + \frac{r_{e}}{r^{2}}\right)dt^{2} 
   + \left(1 - \frac{2MG_{N}}{r} + \frac{r_{e}}{r^{2}}\right)^{-1}dr^{2} + r^{2}d\Omega^{2}_{S^{2}}\nonumber\\
   \\
   A &=& \frac{q}{2}\cos\theta\,d\phi\,,\quad r_{e}^{2} = \frac{\pi q^{2}G_{N}}{g^{2}}\,.\nonumber 
\end{eqnarray}
In the presence of the magnetic field, the spectrum of a charged fermion exhibits a Landau-level structure which receives two contributions; an orbital contribution and a magnetic dipole contribution. This structure turns out to be key to the existence and stability of 4-dimensional traversable wormholes \cite{Maldacena:2018gjk}. 
\\

\noindent
This article is a continuation of our results reported in \cite{Murugan:2018hsd} in which, in Section 2, we expand on our results for the case of a spinless quantum particle confined to the dipole sphere and, in Section 3, substantially extend our study to treat a spin-$\frac{1}{2}$ particle. The latter is of particular relevance to the phenomenology of the fullerine and fullerite configurations (at least in the continuum limit) above and this will be the kind of systems we have in mind for the bulk of this work. We conclude in Section 4 with a discussion of some implications and applications of our results.

\section{Spinless particle on a dipole sphere}
\label{sec:spin0}

Haldane's monopole sphere shares many similarities with the more familiar problem of a particle moving on a plane in a perpendicular magnetic field. In particular, in the limit where the radius of the sphere $R\to\infty$, the two problems converge. For the dipole system that will be the subject of interest here, this is clearly no longer the case since the magnetic field experienced by the particle is no longer uniform. Since the dipole breaks the $O(3)$ symmetry of the Haldane problem to a $U(1)$ about the dipole axis, in the large radius limit, and sufficiently close to the poles of the sphere, the system reduces to the planar Landau problem but in an {\it inhomogeneous} magnetic field with azimuthal  symmetry.\\

\subsection{Oblate spheroidal coordinates}

\begin{figure}[H]
    \centering
	\includegraphics[scale=0.25]{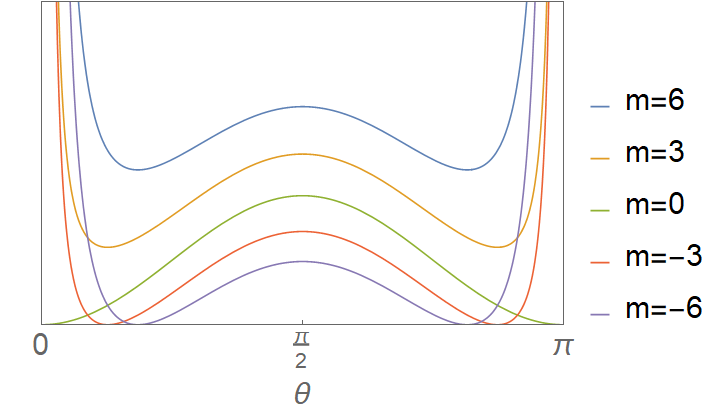} 
    \caption{The effective potential seen by a spinless particle interacting with the dipole field plotted over a polar angle range from the north pole ($\theta=0$) to the south pole ($\theta = \pi$), for various values of the angular momentum quantum number $m$ and magnetic `charge' $Q=20$.}\label{effpot}
\end{figure}

\noindent
In units of $c=1$, the Hamiltonian for the system is given by
\begin{align}
 H &= \frac{1}{2M}(-i\hbar\bm{\nabla}-e\bm{A})^2 \nonumber  \\
     &=  \frac{\hbar^{2}}{2MR^{2}}\!\!\left[-\frac{1}{\sin\theta}\frac{\partial}{\partial\theta}\sin\theta
   \frac{\partial}{\partial\theta} + \left(\frac{i}{\sin\theta} \frac{\partial}{\partial\phi}
   - Q\sin\theta\right)^{2}\right] \,,
\end{align}
where, to facilitate comparison with the spherical monopole system, we have defined 
$Q\equiv e|\bm{\mu}|/\hbar R$. Substituting this into the Schr\"odinger equation and introducing the separable ansatz $\Psi^{m}_{l}(\theta,\phi) = e^{im\phi}\psi_{l}(\theta)$ yields the eigenvalue problem
\begin{equation}\label{dipeval}
   \widetilde{E}_{l,m}\psi_{l} = -\frac{1}{\sin\theta}\frac{\partial}{\partial\theta}\left(\sin\theta
   \frac{\partial\psi_{l}}{\partial\theta}\right) +\frac{\left(m+Q\sin^{2}\theta\right)^{2}}
   {\sin^{2}\theta}\psi_{l}\,,
\end{equation}
where, to avoid confusion between the mass of the particle and the quantum number labelling the eigenfunction, we define 
$\widetilde{E}_{l,m} \equiv 2MR^{2}E_{l,m}/\hbar^{2}$. Before solving the eigenvalue problem, let's build some intuition for what to expect. Using (\ref{dipeval}) the Hamiltonian can be written as the sum of a kinetic and a potential term
\begin{equation}
   H = -\frac{1}{\sin\theta}\frac{\partial}{\partial\theta}\left(\sin\theta
   \frac{\partial}{\partial\theta}\right) + V(\theta;m,Q), \quad \text{where}\quad V(\theta;m,Q)=\frac{\left(m+Q\sin^{2}\theta\right)^{2}}
   {\sin^{2}\theta} \,.
\end{equation}
Many of the features of the quantum states can be deduced already from the potential. In particular, the potential is positive, symmetric about the equator ($\theta = \pi/2$) and has real zeros at 
\begin{eqnarray}
   \theta_{\pm} = \sin^{-1}\left(\pm\sqrt{-\frac{m}{Q}}\right)\,,
\end{eqnarray}
only for momenta in the range $-Q\leq m \leq 0$. For momenta in the range $0<m<Q$, the potential retains the double-well profile but has no real zeros, while for $|m|>Q$ it is a single-well. Physically, this can be understood as follows: since the sign of the cross-term in the potential depends only on the sign of the momentum, only negative momentum states can decrease their energy by coupling to the magnetic field. At fixed $Q$, as $|m|$ is increased, the minima of such states move away from the poles accompanied by a decrease in the height of the local maximum - see Figure \ref{effpot}. Consequently, we expect that states with larger momentum will localise closer to the equator. Conversely, fixing $|m|$ below its critical value and increasing the strength of the magnetic field (and consequently $Q$) pushes out the minima of the effective potential to the poles, localising more states near the polar regions.

The Schr\"odinger equation (\ref{dipeval}) admits a solution in terms of known functions - the analog of spherical harmonics for an oblate spheroid. To see this, let's change variables to $z = \cos\theta$, and write the differential equation in the form
\begin{align}\label{sph}
    \frac{d}{dz}\left[\left(1-z^{2}\right)\frac{d}{dz}\right]\psi 
    + \left[\lambda_l^m+Q^{2}z^{2}-\frac{m^{2}}{(1-z^{2})}\right]\psi=0,\,\,
\end{align}
where 
\begin{align}\label{Dspec}
\lambda_{l}^m \equiv \widetilde{E}- 2mQ - Q^{2}\,.
\end{align}
This can be recognized as an {\it angular oblate spheroidal differential equation} \cite{Morse:1953}. Its solutions are the angular oblate spheroidal wavefunctions, 
\begin{align}\label{sphew}
S_{l}^m(z;Q) &= \sum_{n}d_{n}^{l,m}(Q)P^m_{n}(z)\,, \qquad l-|m| = 0,1,2,\ldots
\end{align}
where $P_n^m(z)$ are the associated Legendre functions, and the coefficient functions $d_{n}^{l,m}(Q)$ satisfy a three-term recursion relation 
\cite{Morse:1953}
\begin{eqnarray}
   \alpha_{n}d_{n+2} + \left(\beta_{n}-\lambda_{l}^m\right)d_{n} + \gamma_{n}d_{n-2} = 0\,,
\end{eqnarray}
whose coefficients 
\begin{align}
\alpha_{n} &= - \frac{(n+2m+1)(n+2m+2)}{(2n+2m+3)(2n+2m+5)}Q^{2}\,, \nonumber \\
\beta_{n} &=(n+m)(n+m+1) -  Q^{2}\frac{2(n+m)(n+m+1)-m^{2}-1}
   {(2n+2m+3)(2n+2m-1)}\,,\nonumber\\
\gamma_{n} &= \frac{n(1-n)Q^{2}}{(2n+2m-1)(2n+2m-3)}\,,
\end{align}
are functions of the momentum $m$, and $Q$. The recursion relation itself may be solved by, for example, the method of continued fractions but this will not be necessary for our purposes since many of the properties of the solution can be inferred from its series expansion.\\

\noindent
The angular spheroidal wave equation \eqref{sph} has two regular singular points at $z= 1$ and $z=-1$, corresponding to the north and south poles of the sphere respectively. Using the fact that in the interval $z\in[-1,+1]$, the associated Legendre functions can be expressed as derivatives of Legendre polynomials of the first kind, $P^{m}_{n}(z) = (1-z^{2})^{m/2}d^{m}P_{n}(z)/dz^{m}$, the spheroidal wavefunctions (\ref{sphew}) can be written as
\begin{eqnarray}\label{sphfun}
    S_l^m(z;Q) =  (1-z^{2})^{m/2}\sum_{n}d_{n}^{l,m}(Q)\frac{d^{m}
    P_{n}(z)}{dz^{m}}\,,
\end{eqnarray} 
In general, solutions that are finite at $z=\pm 1$ will diverge at $z=\mp 1$. However, for a discrete set of eigenvalues $\lambda^m_l$, the series will converge to solutions that are finite at both poles. These eigenvalues are fixed by the requirements that the wavefunction remain finite at $z=\pm1$, and this normalisability condition quantizes the energy spectrum via the relation in (\ref{Dspec}). \\

\begin{figure}[H]
    \begin{subfigure}{0.3\textwidth}
    \centering
	\includegraphics[scale=0.22]{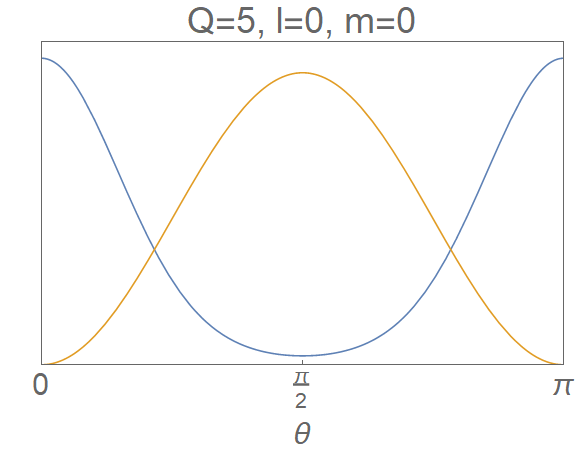} 
    \end{subfigure}
    \begin{subfigure}{0.3\textwidth}
    \centering
    \includegraphics[scale=0.22]{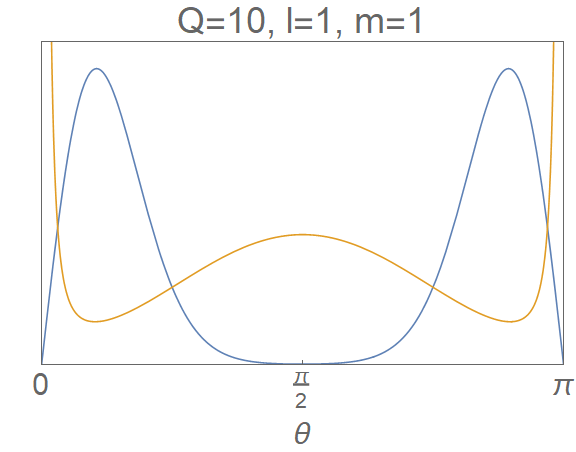} 
    \end{subfigure}
	\begin{subfigure}{0.3\textwidth}
    \centering
    \includegraphics[scale=0.22]{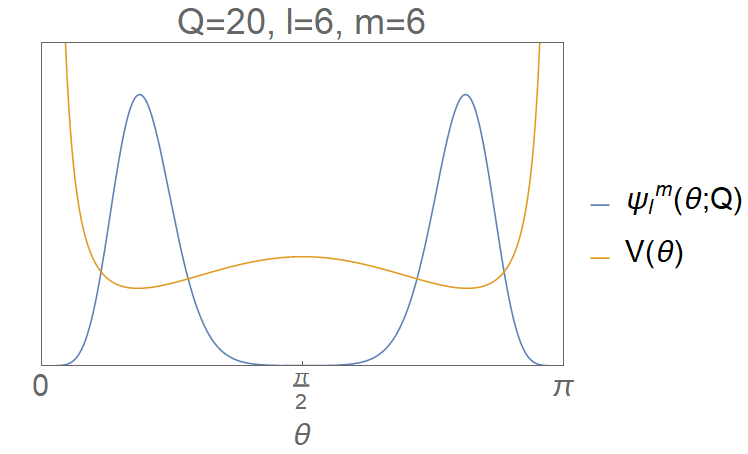} 
    \end{subfigure}
    \caption{Wavefunction $\psi^m_l(\theta;Q)$ and potential $V(\theta)$ for various low energy states. As expected, the wavefunctions are localized at the minima of the effective potential.}\label{effpot2}
\end{figure}

\noindent
The spheroidal eigenvalues are real-valued and satisfy the conjugation relation $\lambda_{l}^m = \lambda^{-m}_l$ and the equality $\lambda^{m}_l<\lambda_{l+1}^{m}$. This ordering means that for a given value of $m$, the smallest value of the eigenvalue is that for which $l=|m|$. Furthermore, for fixed $m$, the corresponding set of eigenfunctions with different $l$ values are mutually orthogonal. Consequently, the full wavefunctions are orthonormal with respect to both quantum numbers,
\begin{align}
\bra{\Psi^{m}_l}\ket{\Psi_{l'}^{m'}} = \delta_{l,l'}\delta_{m,m'}\,.
\end{align}
There are a number of normalisation schemes for the angular oblate functions. In the Stratton-Morse scheme which will be most convenient for our purposes, $S_l^m$ can be normalised by imposing that, near $z=1$, it behaves like the associated Legendre functions $P^{m}_{l}$ for all values of $Q$. This in turn requires that the expansion coefficients satisfy
    \begin{eqnarray}
       \widetilde{\sum_{n}}\,\frac{(n+2m)!}{n!}d_{n}^{l,m}(Q) = \frac{(l+m)!}{(l-m)!}\,.
    \end{eqnarray}
    The tilde over the summation sign is an instruction to include only even values of $n$ if $(l-m)\in 2\mathbb{Z}$ and only odd values of $n$ if $(l-m)\in 2\mathbb{Z}+1$. With this, the normalisation constants are given by
    \begin{align}
    \mathcal{N}^{-1} &=\int_{-1}^{1}\Bigl(S_{l}^m(Q,z)\Bigr)^{2}\,dz
    = \widetilde{\sum_{n}}\,\left(d_{n}^{l,m}(Q)\right)^{2}\left(\frac{2}{2n+2m+1}\right)
    \left(\frac{(n+2m)!}{n!}\right).
    \end{align}
Drawing this all together, the normalised single-particle eigenstates are given by
\begin{eqnarray}\label{single}
   \Psi^m_l(\theta,\phi;Q) = \mathcal{N}e^{im\phi}(\sin\theta)^m\,
    \widetilde{\sum_{n}} d_{n}^{l,m}(Q)\,\frac{d^{m}
    P_{n}(\cos\theta)}{d(\cos\theta)^{m}} 
    = e^{im\phi}\psi^m_l(\theta;Q)\,, 
\end{eqnarray}
with $m\in\mathbb{Z}$ and $l = |m|, |m|+1,\ldots$ Note that since the differential equation (\ref{sph}) is invariant under $m\to-m$, so too are the $\psi^m_l(\theta;Q)$. It is the full states $\Psi^m_l(\theta,\phi;Q)$ which carry the information about the sign of $m$ in the phase factor $e^{im\phi}$, and differ in energy depending on this sign. By way of a sanity-check of our deductions based on the effective potential seen by the charged particle, we plot in Figure \ref{effpot2} sample wavefunctions confirming the localization of the wavefunctions as a function of the polar distance from the north pole.


    
\subsection{\texorpdfstring{The Large $Q$ Limit}{}}
Now let's consider some relevant limits of these solutions. First, note that in the case of a free particle ($Q=0$) our solutions reduce to the standard spherical harmonics. Analytic results can also be obtained in the limit $Q\ll1$, and near the north pole ($\theta\rightarrow0$) \cite{Murugan:2018hsd}. However, for the purposes of this article, we will focus on the large $Q$ limit, $Q\gg1$: this corresponds to taking the radius of the sphere $R$ to be small, or dialling up the dipole field strength by increasing $|\bm{\mu}|$. We begin by considering the case $|m|\ll Q$, where we will see the emergence of a Landau level structure. We will then consider the spectrum for more general $|m|$, for which the Landau level picture will be shown to break down, with some new qualitative features emerging in its place.\\ 



\begin{figure}[H]
    \begin{subfigure}{0.5\textwidth}
    \centering
	\includegraphics[scale=0.28]{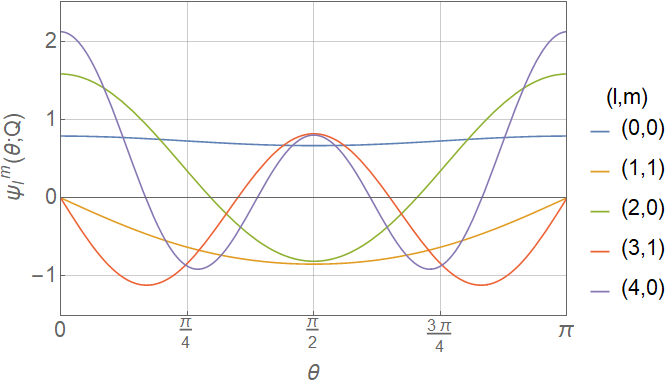} 
    \caption{$Q=1$}\label{Ds1} 
    \end{subfigure}
    \begin{subfigure}{0.5\textwidth}
    \centering
    \includegraphics[scale=0.28]{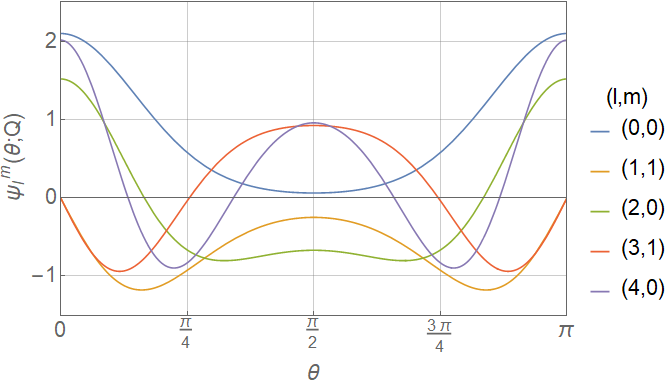} 
    \caption{$Q=5$}\label{Ds2}
    \end{subfigure}
	\\
	\\
	
	\begin{subfigure}{0.5\textwidth}
    \centering
    \includegraphics[scale=0.28]{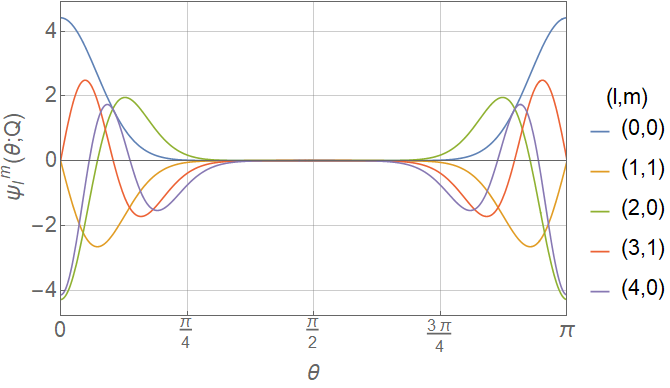} 
    \caption{$Q=20$}\label{Ds3}
    \end{subfigure}
    \begin{subfigure}{0.5\textwidth}
    \centering
    \includegraphics[scale=0.28]{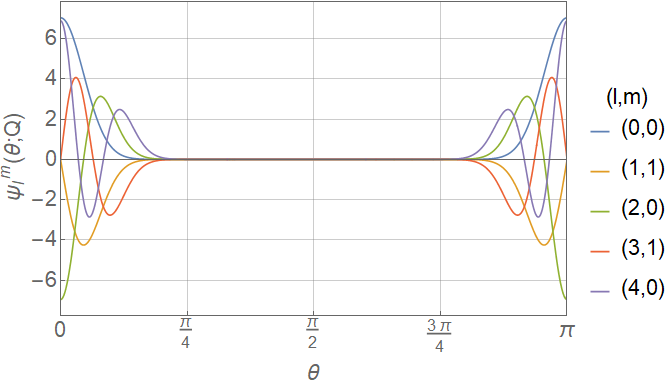} 
    \caption{$Q=50$}\label{Ds4}
    \end{subfigure}
    \caption{$\psi^m_l(\theta;Q)$ for various $l,m$ and $Q$.}\label{DDs1}
\end{figure}

\noindent
We plot in Figure \ref{DDs1} a number of representative states for varying values of $Q$, to observe the behaviour of the system as it approaches the large $Q$ limit. For $Q=1$ the particle states are relatively delocalised over the entire sphere. As the field strength is increased (Figure \ref{DDs1}(b)-(d)), the states become progressively more localised around the poles with negligible amplitude in an expanding region surrounding the equator. This localisation is stronger for states with lower values of the combination of quantum numbers $l-m$. Given then, that in the large $Q$ limit the $|m|\ll Q$ states confine to a region of approximately uniform magnetic field, we anticipate the energy spectrum to exhibit an approximate Landau level structure there. 
\\

\noindent
To confirm this, note that in the limit $|m|\ll Q$ \cite{Berti:2005gp,Hod:2015cqa}, the spheroidal eigenvalues become
\begin{equation}
\lambda_l^m = 2\big[l+1 - \mathrm{mod}(l-m, 2)\big]Q + \mathcal{O}(1)\,.
\end{equation}
If $m$ is integer valued, it follows that $\mathrm{mod}(l-m, 2)=\mathrm{mod}(l+m, 2)$. Substituting this into our expression for the energy spectrum (\ref{Dspec}), allows for the spectrum to be expressed as 
\begin{align}\label{Dstrongspec}
\widetilde{E}_{Q,l,m} = Q^2 + 2 \big[l+m+1 - \mathrm{mod}(l+m, 2)\big]Q + \mathcal{O}(1)\,,
\end{align}
in the large $Q$ limit. This result exhibits a number of noteworthy features.
First; the spectrum only depends on the quantum numbers $l$ and $m$ in the combination $l+m$. The large $Q$ limit spectrum therefore exhibits an (approximate) Landau level structure, where all states of the same $l+m$ are nearly degenerate. Moreover, these (approximate) Landau levels are evenly spaced, with spacing $2Q$. Next, the energies of states with pairwise adjacent $l+m$ values become degenerate in the large $Q$ limit. Finally, the lowest Landau level is comprised of all states with $m\leq0$ that satisfy $l=|m|$ or $l=|m|+1$. We expect this Landau level structure to break down for states with $|m|\gtrsim Q$.\\

\begin{figure}[H]
    \begin{subfigure}{0.5\textwidth}
    \centering
	\includegraphics[scale=0.28]{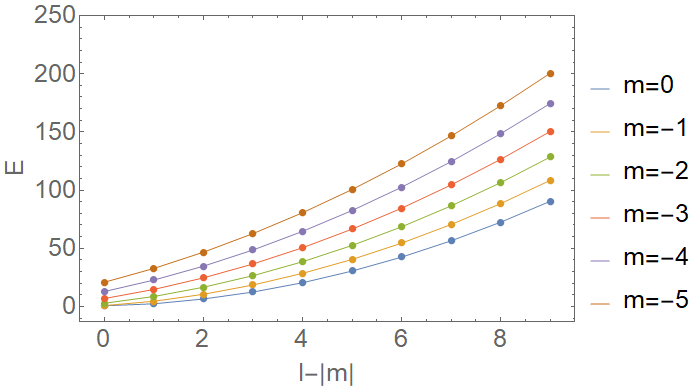} \caption{$Q=1, m\leq0$}\label{zEpMnQ1}
    \end{subfigure} 
    \begin{subfigure}{0.5\textwidth}
    \centering
	\includegraphics[scale=0.28]{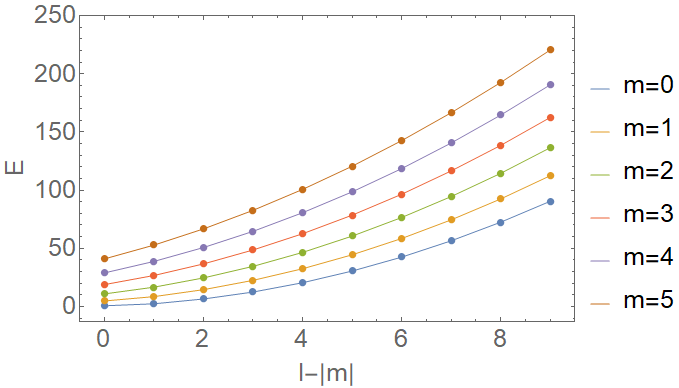} \caption{$Q=1, m\geq0$}\label{zEpMpQ1}
    \end{subfigure} 
    \\ 
    
    \begin{subfigure}{0.5\textwidth}
    \centering
    \includegraphics[scale=0.28]{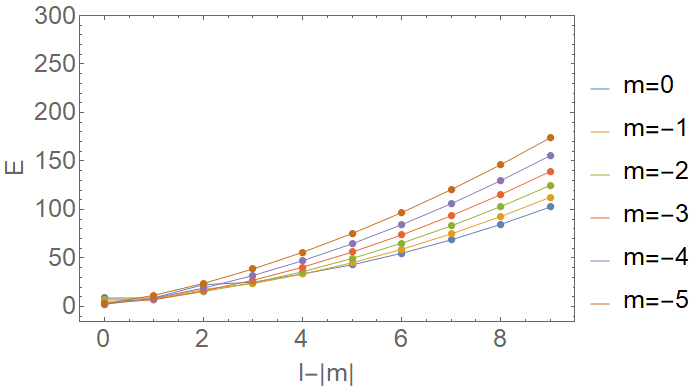} 
    \caption{$Q=5, m\leq0$}\label{zEpMnQ10}
    \end{subfigure}
    \begin{subfigure}{0.5\textwidth}
    \centering
    \includegraphics[scale=0.28]{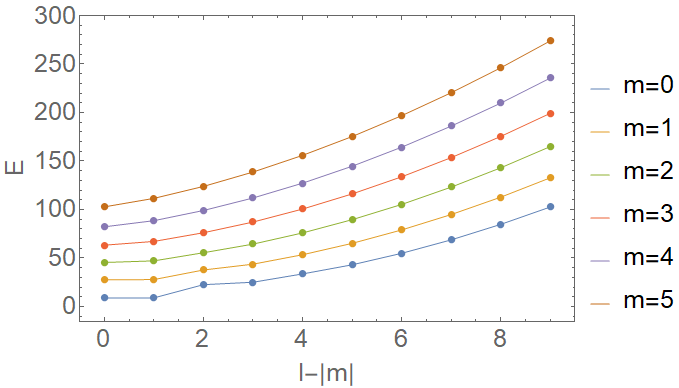} 
    \caption{$Q=5, m\geq0$}\label{zEpMpQ10}
    \end{subfigure}
    \\

    \begin{subfigure}{0.5\textwidth}
    \centering
	\includegraphics[scale=0.28]{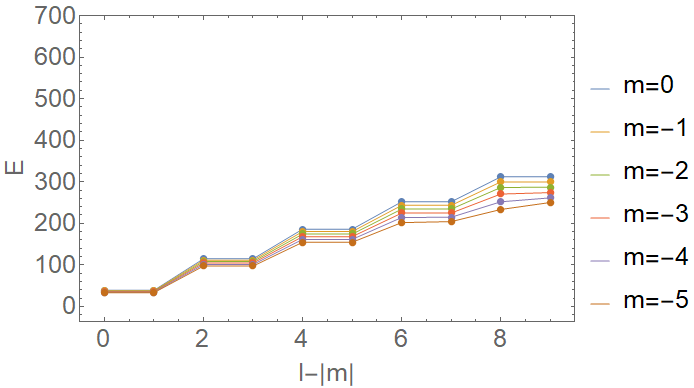} 
    \caption{$Q=20, m\leq0$}\label{zEpMnQ20} 
    \end{subfigure} 
     \begin{subfigure}{0.5\textwidth}
    \centering
	\includegraphics[scale=0.28]{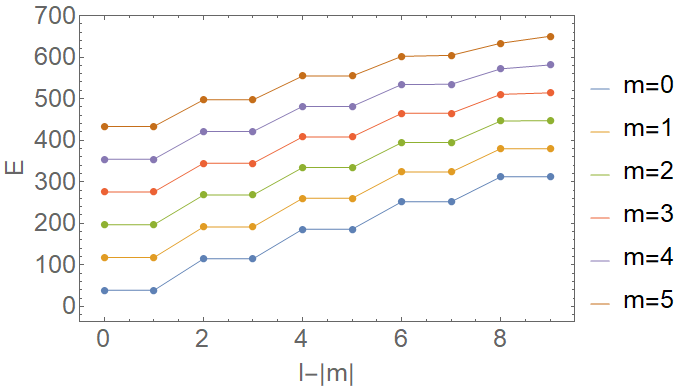} 
    \caption{$Q=20, m\geq0$}\label{zEpMpQ20} 
    \end{subfigure} 
    \\
 
    \begin{subfigure}{0.5\textwidth}
    \centering
    \includegraphics[scale=0.28]{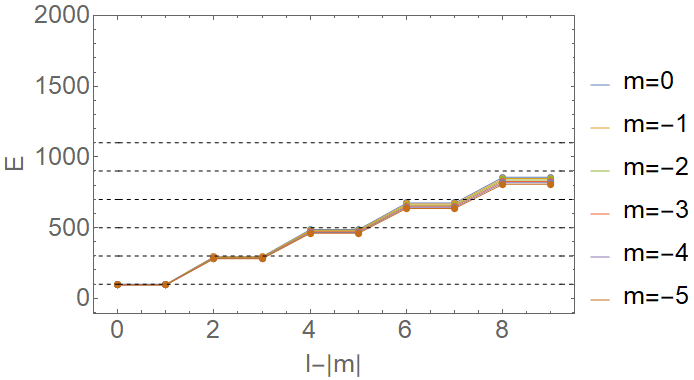} 
    \caption{$Q=50, m\leq0$}\label{zEpMnQ50}
    \end{subfigure}
    \begin{subfigure}{0.5\textwidth}
    \centering
    \includegraphics[scale=0.28]{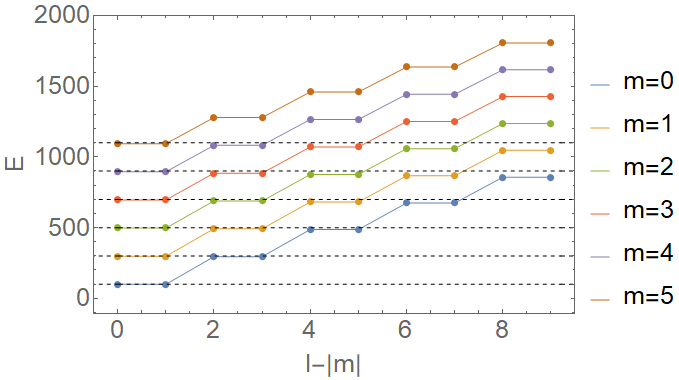} 
    \caption{$Q=50, m\geq0$}\label{zEpMpQ50}
    \end{subfigure}
    
    \caption{Energy spectra for integer $m$.}\label{HEpz}
\end{figure}

\noindent
The spectrum is calculated numerically for states of various $l$ and $m$ and plotted as a function of $l-|m|$ in Figure \ref{HEpz}. 
Notice that, as $Q$ increases, the spectrum converges to the Landau level structure given by the large $Q$ result (\ref{Dstrongspec}), which we plot as black dashed lines. Recall that for a given $Q>1$, states with smaller $l-|m|$ are more strongly localised around the poles than states with greater $l-|m|$. Intuitively, larger $l-|m|$ states thus experience a more varied magnetic field. We expect this field inhomogeneity to perturb their energies, breaking the Landau level degeneracy. This is observed in Figures \ref{HEpz}(g) and (h), where we see the Landau level degeneracy breaking as $l-|m|$ increases. 

\begin{figure}[H]
    \begin{subfigure}{0.5\textwidth}
    \centering
	\includegraphics[scale=0.3]{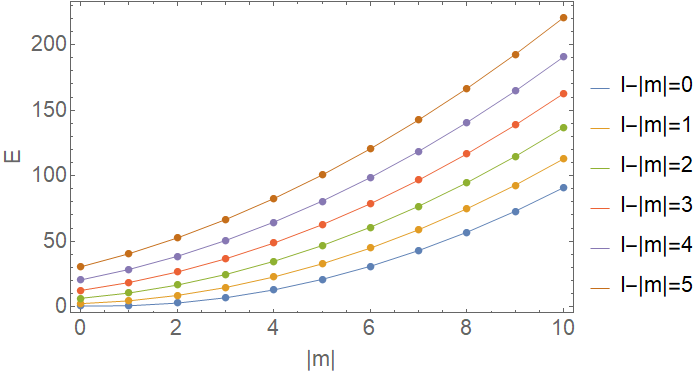} 
    \caption{$Q=1$}\label{largemspec1} 
    \end{subfigure}
    \begin{subfigure}{0.5\textwidth}
    \centering
    \includegraphics[scale=0.3]{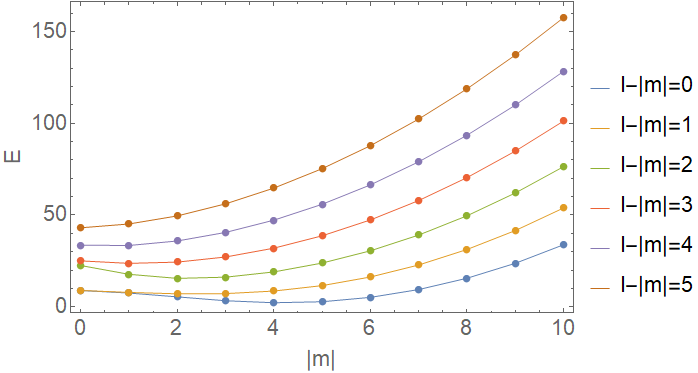} 
    \caption{$Q=5$}\label{largemspec5}
    \end{subfigure}
	\\
	\\
	
	\begin{subfigure}{0.5\textwidth}
    \centering
    \includegraphics[scale=0.3]{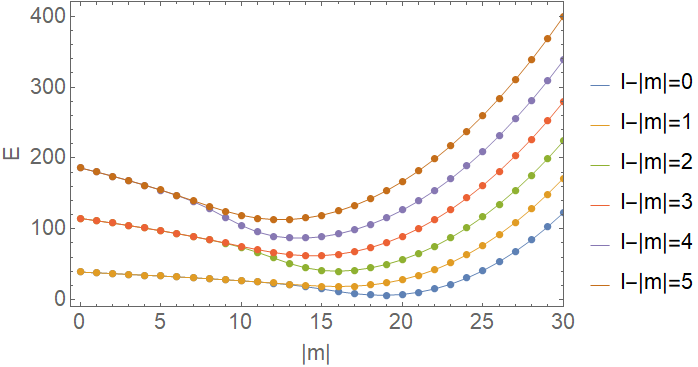} 
    \caption{$Q=20$}\label{largemspec20}
    \end{subfigure}
    \begin{subfigure}{0.5\textwidth}
    \centering
    \includegraphics[scale=0.3]{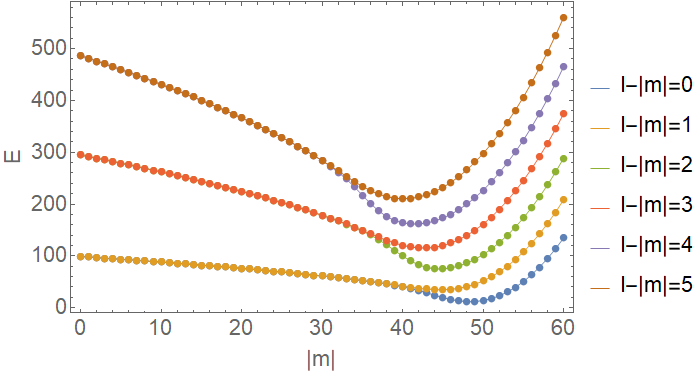} 
    \caption{$Q=50$}\label{largemspec50}
    \end{subfigure}
    \caption{Spectrum for $m\leq0$ and various $Q$.}\label{largemspec}
\end{figure}

\noindent
Now, let's expand the above analysis to include general values of $|m|$. The $m>0$ case is uninteresting; the structure visible in Figure \ref{HEpz}(h) continues in the obvious way for larger $m$\footnote{While the Landau level structure breaks down for larger $m$, the relevant point is that for fixed $l-|m|$, states with larger $m$ have larger energy. This means that the small $m$ Landau level structure is left intact. As we will see, this is no longer the case for $m<0$.}. In Figure \ref{largemspec} we plot the spectrum for $m\leq0$, fixed $l-|m|$ and various values of $Q$. For $Q=20$ and higher, the small $|m|$ Landau level structure of the previous section is visible. However, as $|m|$ increases this degeneracy breaks completely. In particular, for $Q=50$ (see Figure \ref{largemspec}(d)) the energies gradually decrease with increasing $|m|$, until they hit a turning point, after which they begin to rapidly increase. Finally, we note the breaking of the pairwise degeneracy in $l-|m|$ just before the turning point.
\\

\noindent
When $|m|>Q$, the spectrum exhibits a quadratic dependence on $m$. Looking at the Hamiltonian in (\ref{dipeval}) we see that for $|m|$ much larger than $Q$ we can essentially neglect terms proportional to $Q$ in the Hamiltonian, which just leaves  the Hamiltonian for a free particle on the sphere. The energy spectrum of this free particle indeed exhibits a quadratic dependence on its quantum number\footnote{In particular, the free spectrum is $E\sim l(l+1)$ and is independent of $m$.}. In this sense, the dipole potential gives a small $Q$ (relative to $|m|$) correction to the free particle spectrum.

\noindent
One new feature here is a clear ground state for any value of $Q$, whose energy always lies on the $l=|m|$ branch of the spectrum and satisfies $|m|\approx Q$. This ground state is always localised at the equator. 
\\

\begin{figure}[H]
    \begin{subfigure}{0.51\textwidth}
    \centering
    \includegraphics[scale=0.3]{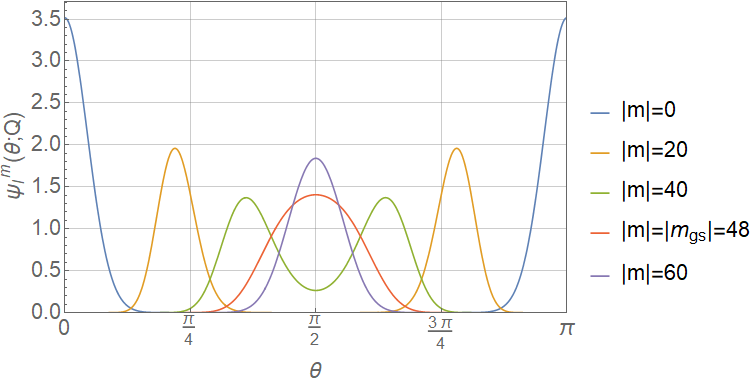} 
    \caption{$l=|m|$.}
    \end{subfigure}
    \begin{subfigure}{0.5\textwidth}
    \centering
    \includegraphics[scale=0.3]{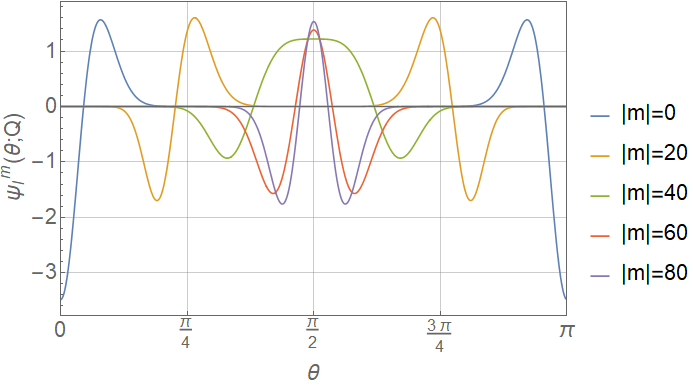} 
    \caption{$l=|m|+2$.}
    \end{subfigure}
    \caption{$\psi^m_l(\theta;Q)$ for $Q=50$ and various $|m|$.}\label{genstates}
\end{figure}

\noindent
We can now describe the behaviour of the quantum states in the large $Q$ limit for general quantum numbers, as illustrated in Figure \ref{genstates}(a) for $l=|m|$. The $m=0$ states are always localised at both poles. As $|m|$ increases, the regions of localisation moves away from the poles and inwards towards the equator, while the energy decreases, in keeping with our observations at the level of the effective potential - larger $|m|$ means localisation further from the pole, where the (normal component of the) magnetic field is weaker, so we expect lower energy. Eventually at some $m=m_{gs}$ the two regions of localisation coalesce at the equator, and we obtain the ground state. Increasing $|m|$ further results in a state more sharply localised at the equator, but this localisation comes at a steep energy cost. The high energy states of the system are then those for which the wavefunction is narrowly localised about the equator. States with larger $l$ show the same qualitative behaviour, but with more nodes in the region of localisation, and a correspondingly higher energy - illustrated for $l=|m|+2$ in Figure \ref{genstates}(b). 

\section{\texorpdfstring{Spin-$\tfrac{1}{2}$ particle on the dipole sphere}{}}
\label{sec:spin1/2}

Having treated at length the Schr\"odinger problem, we turn now to the main focus of this article, a Dirac particle on the dipole sphere. This has direct relevance to experimentally testable configurations, for instance the properties of a  $\mathsf{C}_{60}$ fullerine with a magnetic dipole located at its center, or for that matter in the presence of a uniform magnetic field. While the problem of Landau level quantization for a Dirac particle on the Haldane sphere has recently been studied in some depth \cite{Greiter:2018mkw}, to the best of our knowledge, this is the first calculation carried out in the presence of the more physical dipole field\footnote{The physics of a free Dirac particle on a sphere was studied in \cite{Abrikosov:2002jr}, while the problem was considered for some magnetic potentials amenable to supersymmetric methods in \cite{Jakubsky}.}. Following \cite{Greiter:2018mkw} we start with the standard Dirac Hamiltonian in two dimensions (in units of $c=1$),
\begin{equation}\label{Gham}
  H= v \bm{\hat{n}}\cdot \bigg[ \Big( -i\hbar\bm{\nabla}+e\bm{A} \Big) \times \bm{\sigma} \bigg],
\end{equation}
where $\bm{\hat{n}}$, the unit normal to the surface, in this case is simply $\bm{\hat{n}}=\hat{r}$ and $\bm{\sigma} = \left(\sigma^{x}, \sigma^{y}, \sigma^{z}\right)^{T}$ is a vector of Pauli matrices. In spherical coordinates, the gradient operator is
\begin{equation}\label{nab}
\bm{\nabla}= \frac{1}{R}\bigg( \hat{\theta}\partial_{\theta} + \hat{\phi}\frac{\partial_{\phi}}{\sin\theta} \bigg).
\end{equation}
This in turn means that the Dirac operator $-i\hbar\bm{\nabla}$ is not Hermitian with respect to the natural inner product between two states on the sphere 
\begin{equation}
    \bra{\psi}\ket{\varphi} \equiv R^2\int d\theta d\phi\sin\theta\overline{\psi}\varphi,
\end{equation}
since
 \begin{eqnarray*}
   \bra{\psi}\ket{\partial_{\theta}\varphi}
   =R^2\int d\theta d\phi\sin\theta\overline{\psi}\partial_{\theta}\varphi 
   =-R^2\int d\theta d\phi \sin\theta(\cot\theta\overline{\psi} +\partial_{\theta}\overline{\psi})\varphi,
\end{eqnarray*}
 implies that $\partial_{\theta}^{\dagger}= -(\partial_{\theta}+\cot\theta)$. Similarly, $\partial_{\phi}^{\dagger}= -\partial_{\phi}$. To rectify this, we need to replace this operator in (\ref{Gham}) with the manifestly Hermitian combination
\begin{align}
	\frac{\hbar}{2}(-i\bm{\nabla}+(-i\bm{\nabla})^{\dagger}) &=  -\frac{i\hbar}{2R}\bigg(\hat{\theta}\partial_{\theta} + \hat{\phi}\frac{\partial_{\phi}}{\sin\theta} 	\bigg) + \frac{i\hbar}{2R}\bigg( -\hat{\theta}(\partial_{\theta}+\cot\theta) - \hat{\phi}\frac{\partial_{\phi}}{\sin\theta} \bigg)\nonumber\\
	&= -\frac{i\hbar}{R}\bigg( \hat{\theta}\partial_{\theta} + \hat{\phi}\frac{\partial_{\phi}}{\sin\theta} +\frac{\hat{\theta}}{2}\cot\theta \bigg).
\end{align}
In order to study the dipole potential we again choose, 
\begin{equation}
	\bm{A}=A(\theta)\hat{\phi}=\frac{|\bm{\mu}|}{R^{2}}\sin\theta\hat{\phi},
\end{equation}
and using $\bm{\sigma}=\hat{\theta}\sigma_x + \hat{\phi}\sigma_y + \hat{r}\sigma_z$, the Hamiltonian, (\ref{Gham}), then becomes
\begin{align}
	H &= -\frac{i\hbar v}{R} \hat{r}\cdot \bigg[ \hat{\theta}\Big( \partial_{\theta} + \frac{1}{2}\cot\theta\Big)\cross \bm{\sigma} + \hat{\phi} \Big( \frac{\partial_{\phi}}{\sin\theta} + i \frac{eR}{\hbar} A(\theta) \Big)\cross \bm{\sigma} \bigg]\nonumber\\
	&= -\frac{i\hbar v}{R} \bigg[ \Big( \partial_{\theta} + \frac{1}{2}\cot\theta\Big)\sigma_y - \Big( \frac{\partial_{\phi}}{\sin\theta} + i \frac{eR}{\hbar} A(\theta) \Big)\sigma_x \bigg],\nonumber
\end{align}
which in explicit matrix form reads
\begin{align}\label{Gh}
	H &=  \frac{\hbar v}{R} 
	\begin{pmatrix} 
	0 & -\partial_{\theta} + \frac{i\partial_{\phi}}{\sin\theta} -\frac{1}{2}\cot\theta -\frac{eR}{\hbar} A(\theta) \\
	\partial_{\theta} + \frac{i\partial_{\phi}}{\sin\theta} +\frac{1}{2}\cot\theta -\frac{eR}{\hbar} A(\theta) & 0 \\
	\end{pmatrix}.
\end{align}
To solve the eigenvalue problem for the Dirac operator, we write
\begin{align}\label{dipschro}
\begin{pmatrix} 
	0 & -\partial_{\theta} + \frac{i\partial_{\phi}}{\sin\theta} -\tfrac{1}{2}\cot\theta -Q\sin\theta \\
	\partial_{\theta} + \frac{i\partial_{\phi}}{\sin\theta} +\tfrac{1}{2}\cot\theta-Q\sin\theta & 0 \\
	\end{pmatrix}\Psi=\mathcal{E}\Psi,
\end{align}
where $\mathcal{E}\equiv\tfrac{R}{\hbar v}E$, and we have again defined $Q\equiv \frac{e|\bm{\mu}|}{\hbar R}$. The separable ansatz $\Psi=e^{im\phi}\frac{1}{\sqrt{\sin\theta}}\Phi(\theta)$, puts the eigenvalue problem into the form
\begin{align}\label{Hdir}
    \begin{pmatrix}
    0 & H_+ \\
    H_- & 0
    \end{pmatrix} \Phi = \mathcal{E}\Phi,
\end{align}
with
\begin{align}
H_{\pm} &\equiv  \mp \partial_{\theta}-\frac{m}{\sin\theta}- Q \sin\theta.
\end{align}
There are two points to note here. The first is that the operators $H_+$ and $H_-$  map to each other under the transformation $\theta\rightarrow\pi-\theta$ and the second
is that there exists an analytically solvable, zero-energy configuration which can be found in the decoupled case when $\mathcal{E}=0$. This in turn leads to two independent zero-energy solutions given by
\begin{equation}
    \Psi = \begin{pmatrix} \tan(\theta/2)^m e^{-Q\cos\theta} \\ 0 \end{pmatrix}e^{im\phi} \quad\text{and}\quad
    \Psi = \begin{pmatrix} 0 \\ \cot(\theta/2)^m e^{Q\cos\theta} \end{pmatrix}e^{im\phi}\,.
\end{equation}
In fact, these are the {\it only} exact analytic solutions and are only normalizable for $m=0$. To proceed to the case of non-zero $\mathcal{E}$, we define the two component spinor (\cite{Jellal:2007da}) satisfying 
\begin{equation}\label{2sP}
    H^2 \begin{pmatrix} f_1 \\ f_2 \end{pmatrix} = E^2 \begin{pmatrix} f_1 \\ f_2 \end{pmatrix}\,.
\end{equation}
It is readily verified that the associated spinor 
\begin{equation}\label{evecs1}
    \begin{pmatrix} \psi^{\pm}_1 \\ \psi^{\pm}_2 \end{pmatrix} \equiv H \begin{pmatrix} f_1 \\ f_2 \end{pmatrix} \pm E \begin{pmatrix} f_1 \\ f_2 \end{pmatrix}
\end{equation}
is an eigenvector of $H$ with eigenvalue $\pm E$. For the Hamiltonian in (\ref{Hdir}), the condition (\ref{2sP}) yields the decoupled equations
\begin{align}
    H_+H_-f_1 = \mathcal{E}^2f_1\,,\quad H_-H_+f_2 = \mathcal{E}^2f_2\,, \label{decou} 
\end{align}
for the spinor components $f_{1}$ and $f_{2}$. The fact that the operators $H_+H_-$ and $H_-H_+$ map into each other by the transformation $\theta\rightarrow\pi-\theta$, means that the spinor components are related through
\begin{align}\label{so}
f_2(\theta)&=f_1(\pi-\theta)\,.
\end{align}
The prescription (\ref{evecs1}) then yields the following solutions 
\begin{align}
    \psi_{1}^{\pm}(\theta) &= \Big( -\partial_{\theta} -\frac{m}{\sin\theta} - Q \sin\theta\Big)f_1(\pi-\theta) \pm \mathcal{E} f_1(\theta)\,,\label{firstsol} \\  
    \psi_{2}^{\pm}(\theta) &= \Big( \partial_{\theta} -\frac{m}{\sin\theta} - Q \sin\theta  \Big)f_1(\theta) \pm \mathcal{E} f_1(\pi-\theta)\label{secsol}\,.
\end{align}
Note that we only need to calculate (\ref{firstsol}); we can obtain (\ref{secsol}) by applying the transformation $\theta\rightarrow\pi-\theta$. Solving the full eigenvalue problem (\ref{Hdir}) therefore requires finding the solutions $f_m(\theta)$ to the second order decoupled equation (\ref{decou}), 
\begin{align}\label{mnonzero1}
   -\frac{d^{2}f_m}{d\theta^{2}} + \Big( (m \csc\theta +
     Q \sin\theta)^2+Q\cos\theta - m \cot\theta \csc\theta \Big) f_m = \mathcal{E}_m^2 f_m\,. 
\end{align}
Putting this together then, spinor solutions to (\ref{dipschro}) are given by
\begin{align}\label{fullsol}
   \Psi^{\pm}_m(\theta,\phi;Q)= \frac{1}{\sqrt{\sin\theta}}\begin{pmatrix} \psi_{m}^{\pm}(\theta;Q) \\ \psi_{m}^{\pm}(\pi-\theta;Q) \end{pmatrix}e^{im\phi}\,,
\end{align}
where
\begin{align}\label{psi1}
   \psi_m^{\pm}(\theta;Q) = -\Big(\frac{m}{\sin\theta} + Q \sin\theta +\partial_{\theta} \Big)f_m(\pi-\theta) \pm \mathcal{E}_m f_m(\theta)\,,
\end{align}
$f_m$ is a solution to (\ref{mnonzero1}) and $\pm\mathcal{E}_m$ is the energy eigenvalue of the spinor $\Psi^{\pm}_m$, now all labelled by the momentum quantum number $m$.
\\

\noindent
As a first check, consider the special case $m=0$ which reduces (\ref{mnonzero1}) to
\begin{align}\label{mzero}
   f_m''(\theta) = \Big(Q^2\sin^2\theta +Q\cos\theta - \mathcal{E}_m^2\Big) f_m(\theta) \,,
\end{align}
denoting $'\equiv d/d\theta$. This simplifies even further in the limit $\mathcal{E}_m\gg Q$, to
\begin{align}
   f_m''(\theta) \approx - \mathcal{E}_m^2 f_m(\theta) \quad \Rightarrow \quad f_m(\theta) = A \sin(\mathcal{E}_m\theta)+ B\cos(\mathcal{E}_m\theta) \,.
\end{align}
This in turn can be substituted into (\ref{psi1}), yielding
\begin{align}\label{limpsi1}
    \psi_m^{\pm}(\theta) &\approx C\cos(\mathcal{E}_m\theta) +D\sin(\mathcal{E}_m\theta) \,,
\end{align}
where $C$ and $D$ are complex constants (depending on $\mathcal{E}_m,A$ and $B$) whose exact form doesn't concern us. What does matter, is that in order for (\ref{limpsi1}) to satisfy either Neumann or Dirichlet boundary conditions requires that $\mathcal{E}_m \in \mathbb{Z}$. In other words, $m=0$ states with energies  $\mathcal{E}_m\gg Q$ are oscillatory functions with angular frequency equal to their (quantized) energy values. For more general values of $m$, we need to resort to numerical methods, sadly. 
\\

\noindent
In what follows, we will require both spinor components to be normalizable on the sphere as follows:
\begin{align}
   \bra{\Psi_{m}^{\pm}}\ket{\Psi_{m}^{\pm}}
   =R^2\int d\theta d\phi\sin\theta \left|\frac{1}{\sqrt{\sin\theta}}\psi_{m}^{\pm}(\theta)\right|^2 
   \sim \int d\theta |\psi_{m}^{\pm}(\theta)|^2 =1.
\end{align}
Numerically enforcing this condition quantizes the allowed energy eigenvalues $\mathcal{E}_m$. 


\subsection{\texorpdfstring{Half-odd $m$}{}}

We'll start with the case where the momentum quantum number $m\in \mathbb{Z}+\frac{1}{2}$ and show that when, in addition, $|m|\ll Q$ in the large $Q$ limit, we again see the emergence of an approximate Landau level structure. For more general $|m|$, this again breaks down as in the non-relativistic case, revealing some other interesting qualitative spectral features.\\ 

\begin{figure}[H]
    \begin{subfigure}{0.5\textwidth}
    \centering
	\includegraphics[scale=0.3]{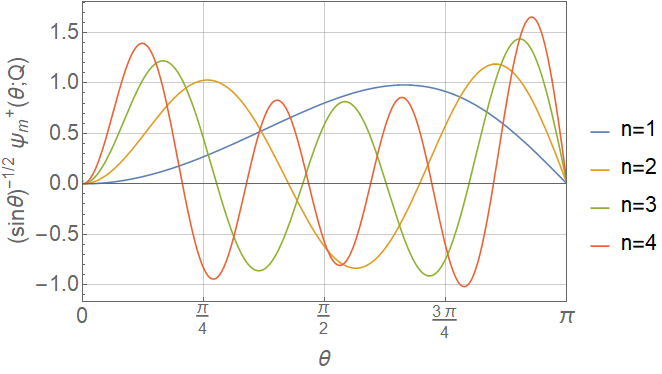} 
    \caption{$Q=1$}\label{DS1} 
    \end{subfigure}
    \begin{subfigure}{0.5\textwidth}
    \centering
    \includegraphics[scale=0.3]{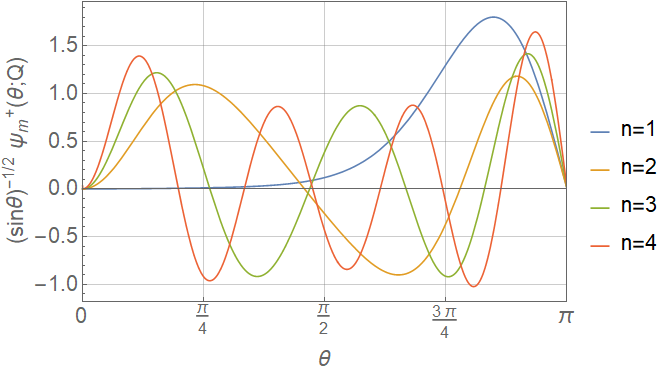} 
    \caption{$Q=5$}\label{DS2}
    \end{subfigure}
	\\
	\\
	
	\begin{subfigure}{0.5\textwidth}
    \centering
    \includegraphics[scale=0.3]{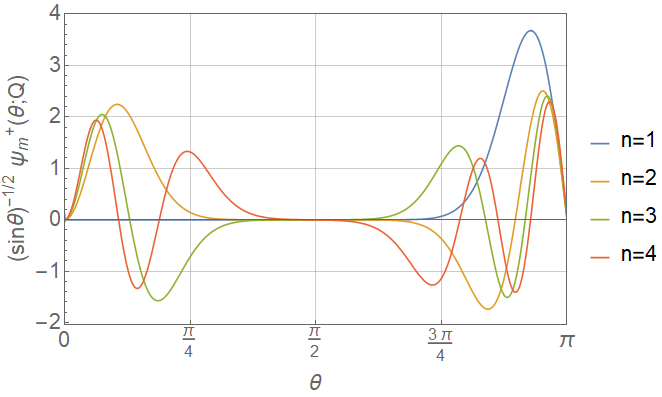} 
    \caption{$Q=20$}\label{DS3}
    \end{subfigure}
    \begin{subfigure}{0.5\textwidth}
    \centering
    \includegraphics[scale=0.3]{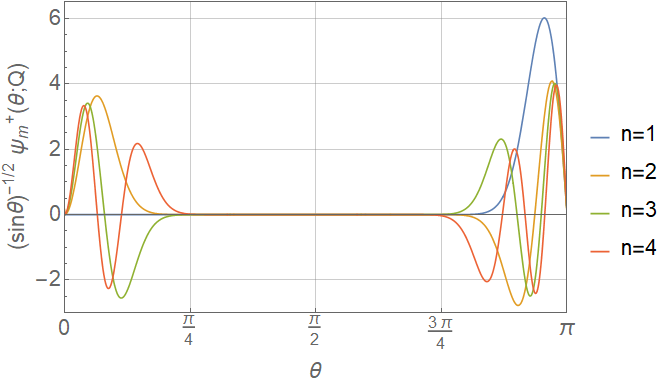} 
    \caption{$Q=50$}\label{DS4}
    \end{subfigure}
    \caption{An illustrative set of positive energy wavefunctions $\frac{1}{\sqrt{\sin\theta}}\psi^+_m(\theta;Q)$ for $m=3/2$ and increasing magnetic field strength $Q$. The quantum number $n$ labels successive allowed energies for a given value of $m$. In each case, we plot the first spinor component, recalling that the second component is obtained by reflecting the wavefunction about the equator.}\label{diracstates}
\end{figure}

\noindent
The states in Figure \ref{diracstates} exhibit the same qualitative behaviour as in the spinless case; as $Q$ is increased they become progressively more localised around the poles with negligible amplitude in an expanding region surrounding the equator. Our earlier argument for the existence of Landau levels still holds and will be verified below.

\begin{figure}[H]
    \centering
    \includegraphics[scale=0.3]{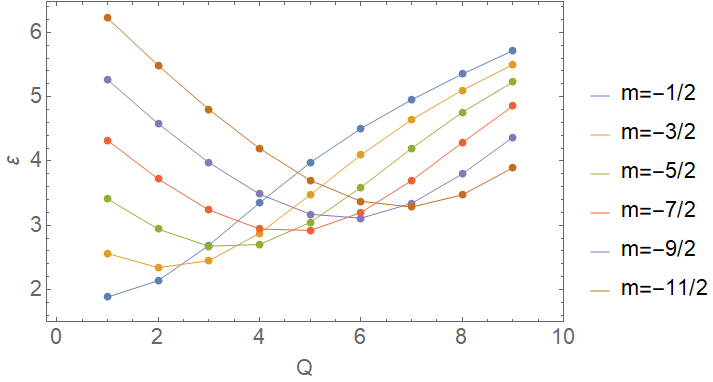} 
    \caption{Level crossing of $n=1$ states as $Q$ is increased.}\label{levelcrossing}
\end{figure}

\begin{figure}[H]
    \begin{subfigure}{0.5\textwidth}
    \centering
	\includegraphics[scale=0.25]{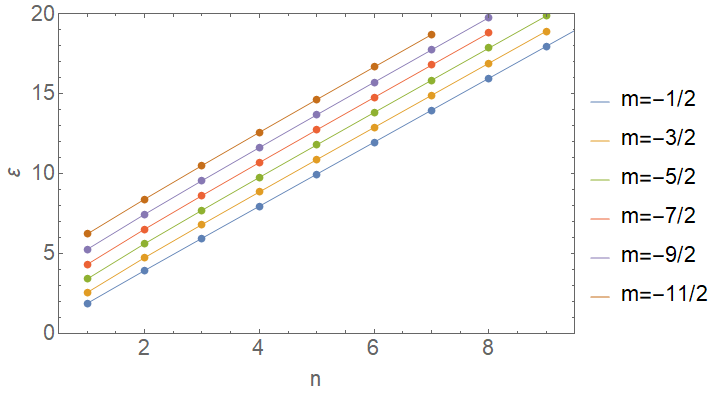}    
	\caption{$Q=1, m<0$}\label{HEpMnQ1}
    \end{subfigure} 
    \begin{subfigure}{0.5\textwidth}
    \centering
	\includegraphics[scale=0.25]{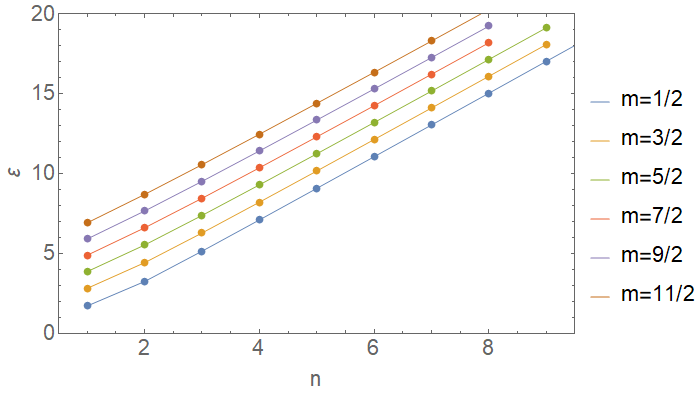}    
	\caption{$Q=1, m>0$}\label{HEpMpQ1}
    \end{subfigure} 
    \\ 
	\\

    \begin{subfigure}{0.5\textwidth}
    \centering
    \includegraphics[scale=0.25]{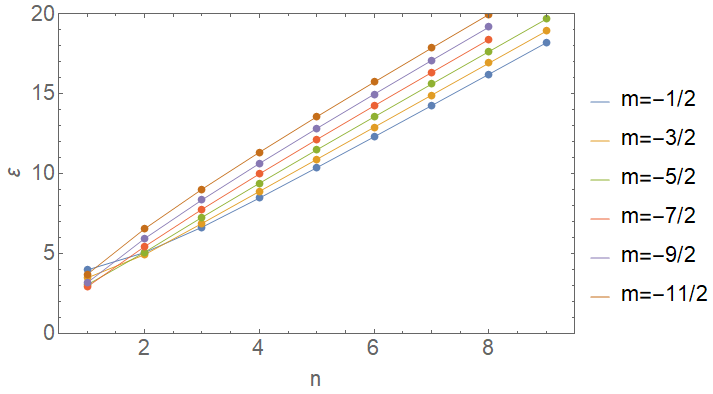} 
    \caption{$Q=5, m<0$}\label{HEpMnQ5}
    \end{subfigure}
    \begin{subfigure}{0.5\textwidth}
    \centering
    \includegraphics[scale=0.25]{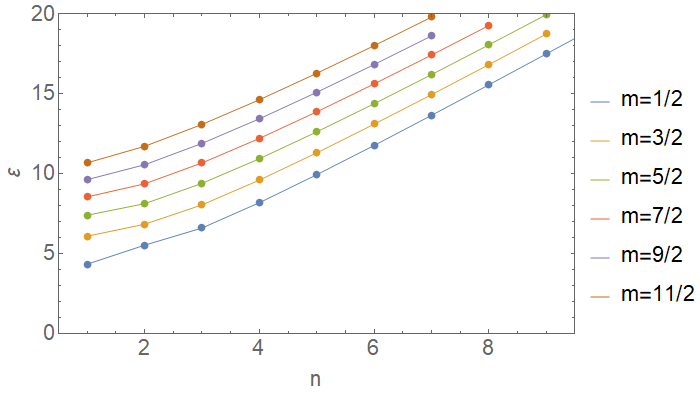} 
    \caption{$Q=5, m>0$}\label{HEpMpQ5}
    \end{subfigure}
    \\
    \\
    
    \begin{subfigure}{0.5\textwidth}
    \centering
	\includegraphics[scale=0.25]{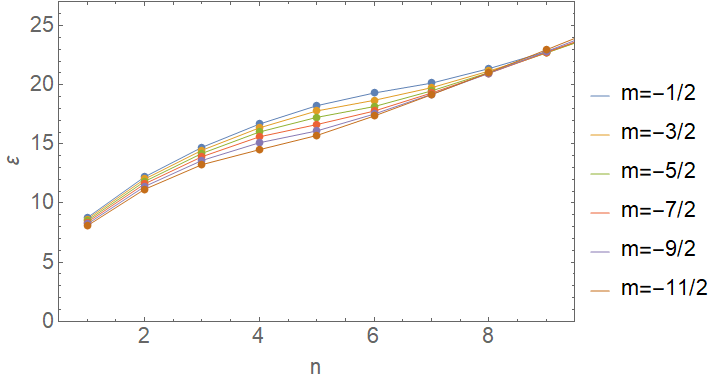} 
    \caption{$Q=20, m<0$}\label{HEpMnQ20} 
    \end{subfigure} 
     \begin{subfigure}{0.5\textwidth}
    \centering
	\includegraphics[scale=0.25]{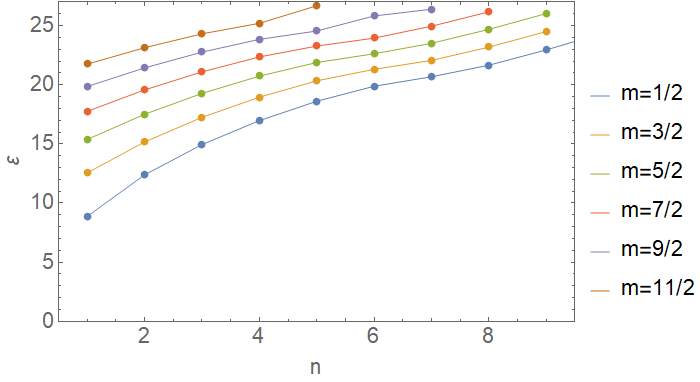} 
    \caption{$Q=20, m>0$}\label{HEpMpQ20} 
    \end{subfigure} 
    \\
    \\

    \begin{subfigure}{0.5\textwidth}
    \centering
    \includegraphics[scale=0.25]{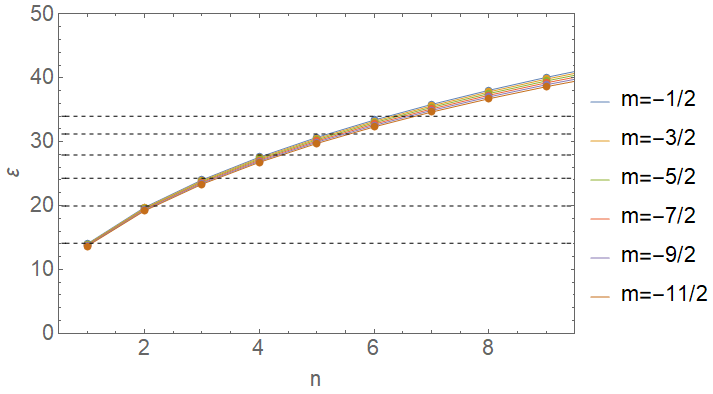} 
    \caption{$Q=50, m<0$}\label{HEpMnQ50}
    \end{subfigure}
    \begin{subfigure}{0.5\textwidth}
    \centering
    \includegraphics[scale=0.25]{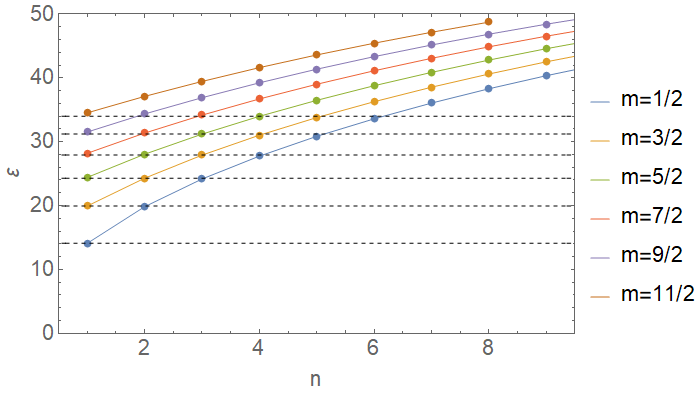} 
    \caption{$Q=50, m>0$}\label{HEpMpQ50}
    \end{subfigure}
    
    \caption{Positive energy spectra for $m$ half-odd.}\label{HEp}
\end{figure}

\begin{figure}[H]
    \begin{subfigure}{0.5\textwidth}
    \centering
	\includegraphics[scale=0.25]{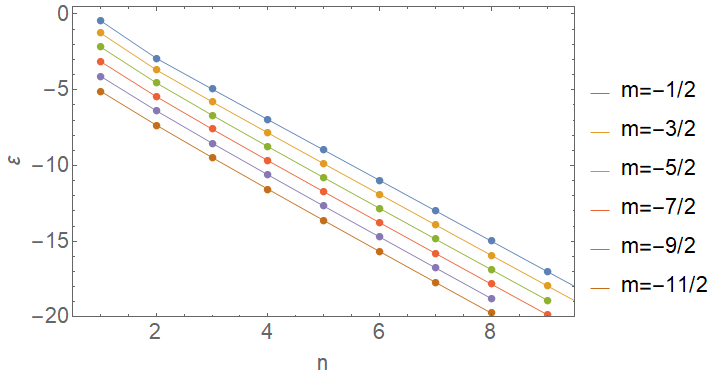}   
	\caption{$Q=1, m<0$}\label{HEnMnQ1}
    \end{subfigure} 
     \begin{subfigure}{0.5\textwidth}
    \centering
	\includegraphics[scale=0.25]{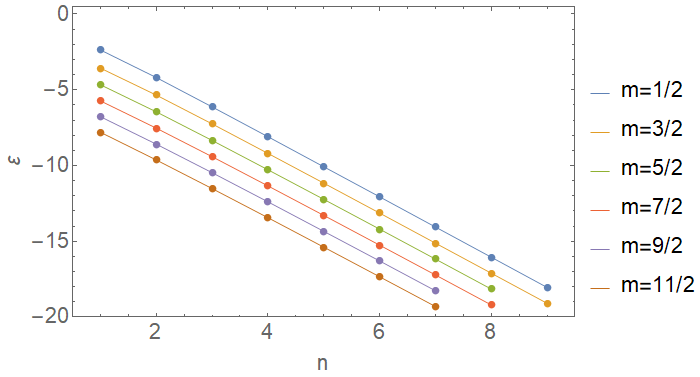}   
	\caption{$Q=1, m>0$}\label{HEnMpQ1}
    \end{subfigure} 
    \\
    \\
    
    \begin{subfigure}{0.5\textwidth}
    \centering
    \includegraphics[scale=0.25]{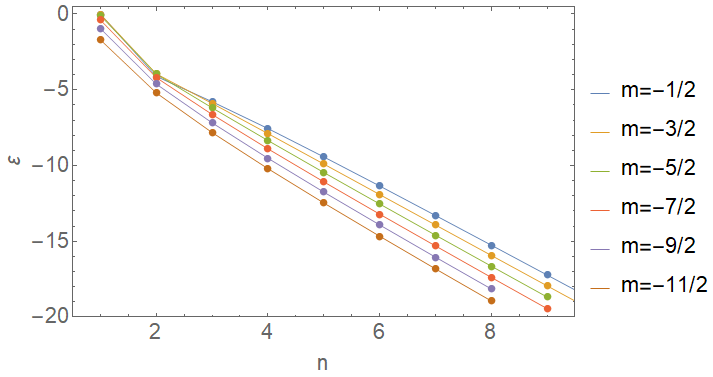} 
    \caption{$Q=5, m<0$}\label{HEnMnQ5}
    \end{subfigure}
    \begin{subfigure}{0.5\textwidth}
    \centering
    \includegraphics[scale=0.25]{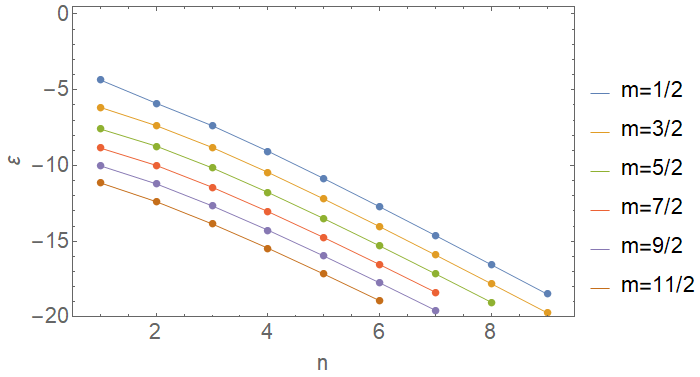} 
    \caption{$Q=5, m>0$}\label{HEnMpQ5}
    \end{subfigure}
    \\
    \\
    
    \begin{subfigure}{0.5\textwidth}
    \centering
	\includegraphics[scale=0.25]{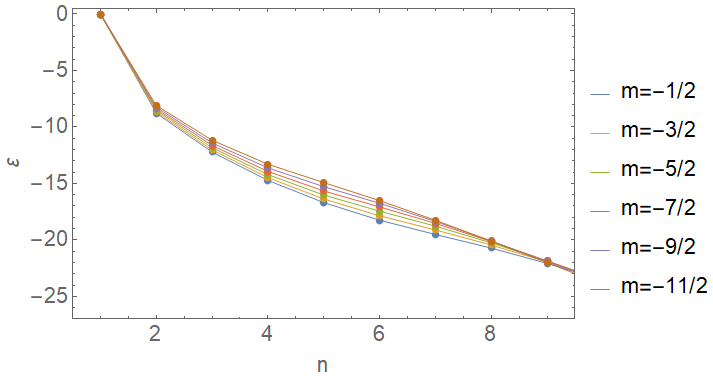} 
    \caption{$Q=20, m<0$}\label{HEnMnQ20} 
    \end{subfigure} 
       \begin{subfigure}{0.5\textwidth}
    \centering
	\includegraphics[scale=0.25]{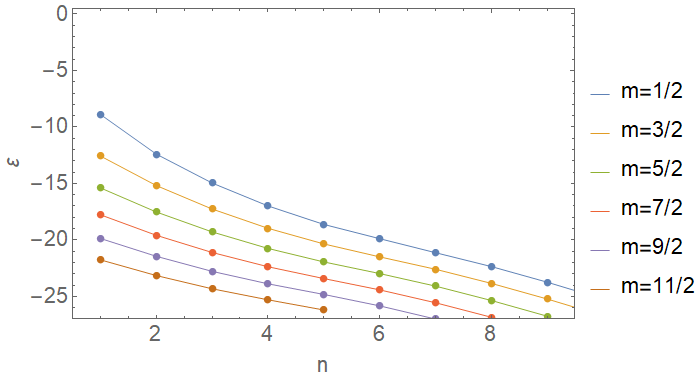} 
    \caption{$Q=20, m>0$}\label{HEnMpQ20} 
    \end{subfigure} 
    \\
    \\
    
    \begin{subfigure}{0.5\textwidth}
    \centering
    \includegraphics[scale=0.25]{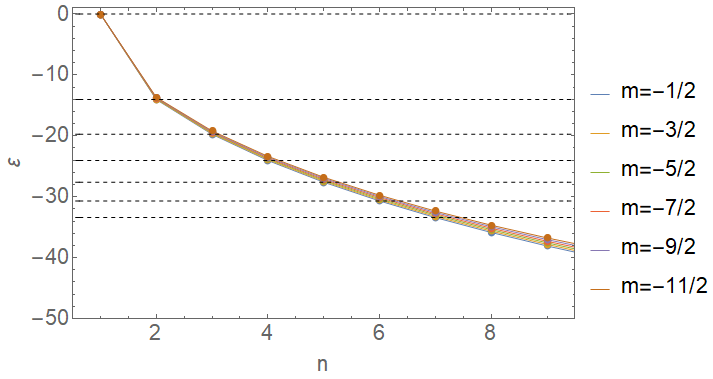} 
    \caption{$Q=50, m<0$}\label{HEnMnQ50}
    \end{subfigure}
    \begin{subfigure}{0.5\textwidth}
    \centering
    \includegraphics[scale=0.25]{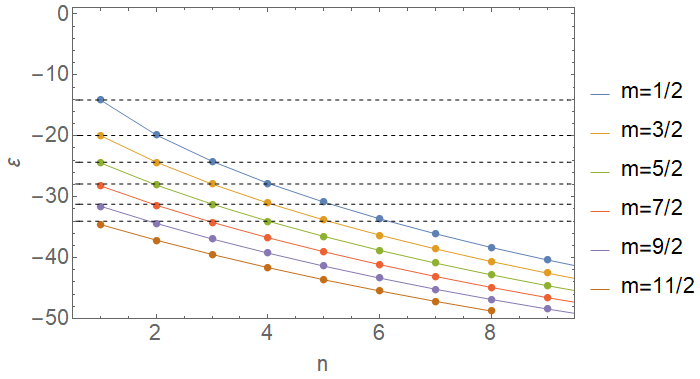} 
    \caption{$Q=50, m>0$}\label{HEnMpQ50}
    \end{subfigure}
    \caption{Negative energy spectra for $m$ half-odd.}\label{HEn}
\end{figure}

\noindent
In Figures \ref{HEp} and \ref{HEn} we compute the positive and negative energy spectra for various values of $m$ and $Q$. For $m>0$, we note that as the dipole strength is increased to $Q=50$ a series of systematic (approximate) degeneracies become visible in the spectrum (see Figures \ref{HEp}(h) and \ref{HEn}(h)); states which share an $m+n$ value have degenerate energies. This is easy to see for the first few "levels" marked in dashed black lines in these Figures. For $Q\gg 50$ our numerical methods become unstable; however, in the $Q\rightarrow\infty$ limit we conjecture that this degeneracy becomes exact for all levels. For $m<0$ on the other hand, a different degeneracy emerges in the large $Q$ limit. In this case, all states with the same $n$ value become degenerate, regardless of $m$ (see Figures \ref{HEp}(g) and \ref{HEn}(g)). This is again clear for the lower levels, after which the degeneracy breaks down at larger $n$ ($n\gtrsim5$). Our analysis again suggests that this degeneracy becomes exact for all levels in the  $Q\rightarrow\infty$ limit. Finally, we note that the $m<0$ and $m>0$ levels coincide. Thus we see the emergence of a Landau level structure in the large $Q$ limit, consistent with our expectation from the non-relativistic study in Section \ref{sec:spin0}. Note that for the positive (negative) spectrum, the spacing between the levels increases with increasing (decreasing) energy. We also note that for small $n$, some level crossing occurs among the various $m\leq0$ states as $Q$ is increased. We show this for  $n=1$ in Figure \ref{levelcrossing}.

\begin{figure}[H]
    \begin{subfigure}{0.5\textwidth}
    \centering
	\includegraphics[scale=0.3]{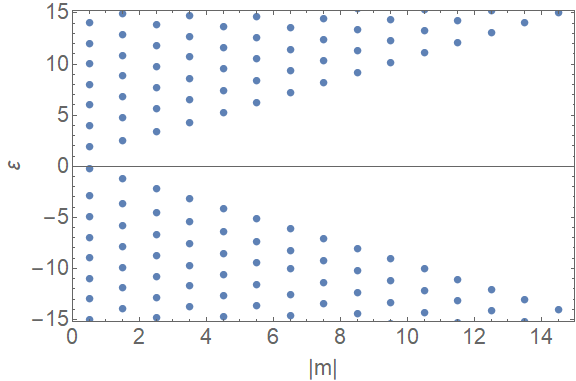} 
    \caption{$Q=1$}\label{Q1variousmDirac} 
    \end{subfigure}
    \begin{subfigure}{0.5\textwidth}
    \centering
    \includegraphics[scale=0.3]{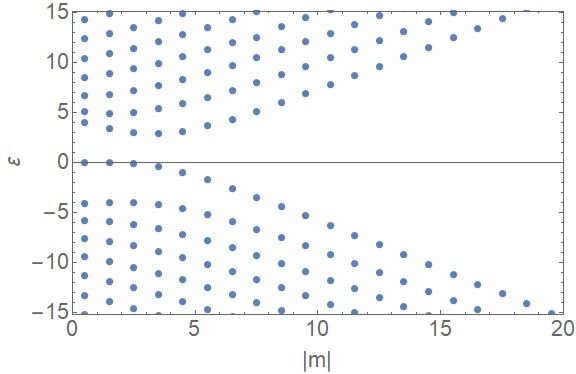} 
    \caption{$Q=5$}\label{Q5variousmDirac}
    \end{subfigure}
	\\
	\\
	
	\begin{subfigure}{0.5\textwidth}
    \centering
    \includegraphics[scale=0.3]{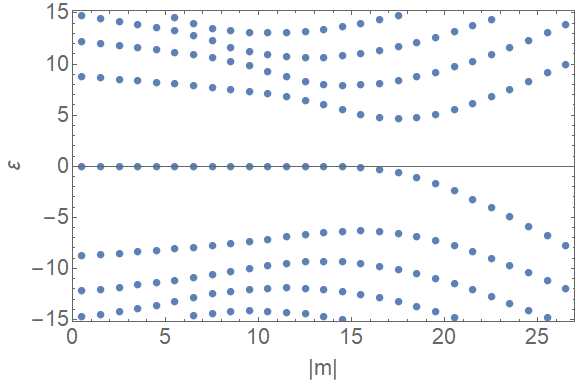} 
    \caption{$Q=20$}\label{Q20variousmDirac}
    \end{subfigure}
    \begin{subfigure}{0.5\textwidth}
    \centering
    \includegraphics[scale=0.3]{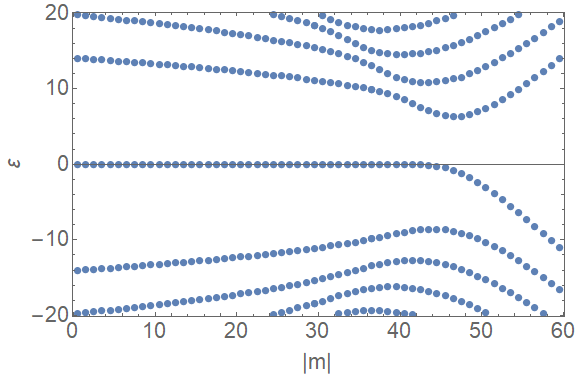} 
    \caption{$Q=50$}\label{Q50variousmDirac}
    \end{subfigure}
    \caption{$m<0$ spectrum for various $Q$.}\label{variousmDirac}
\end{figure}

\noindent
Now let's expand the above analysis to more general values of $|m|$. In Figure \ref{variousmDirac} we plot the spectrum for $m<0$\footnote{As in the spinless system, the general $m>0$ case is not particularly interesting.} for various values of $Q$. The qualitative behaviour of the spectra is entirely analogous to that of the spinless case, except for two new features.\\

\noindent
Recall that in the latter, branches of the spectrum with adjacent $l-|m|$ values were degenerate for small $|m|$, until a Zeeman-like splitting occurs beyond some critical value of $|m|$ (see Figure \ref{largemspec}). Plotting the \textit{absolute value} of all the energies in the spin-$\tfrac{1}{2}$ spectra would produce a similar result. However, each of the pair of degenerate branches of the spin-$\tfrac{1}{2}$ spectrum takes a different sign, effectively lifting the pairwise degeneracy present in the spinless system. This is because one of either $\psi^+_m$ or $\psi^-_m$ in (\ref{psi1}) always vanishes and is thus not a true solution. The alternating manner in which this occurs gives rise to the spectrum in figure \ref{largemspec}, which is not symmetric about $\mathcal{E}=0$ as one might expect. Secondly, the large $Q$ limit now exhibits an (approximately) zero energy "lowest Landau level" (see Figure \ref{variousmDirac}(d)). It is however not a true Landau level, since for sufficiently large values of the momentum quantum number (in this case, $|m|\approx45$) the degeneracy breaks as the energies move away from zero and close the gap to the negative energy band below. However, the system does have a large number of (approximately) degenerate states with energies very close to zero.

\subsection{\texorpdfstring{Integer $m$}{}}

\begin{figure}[H]
    \begin{subfigure}{0.5\textwidth}
    \centering
	\includegraphics[scale=0.25]{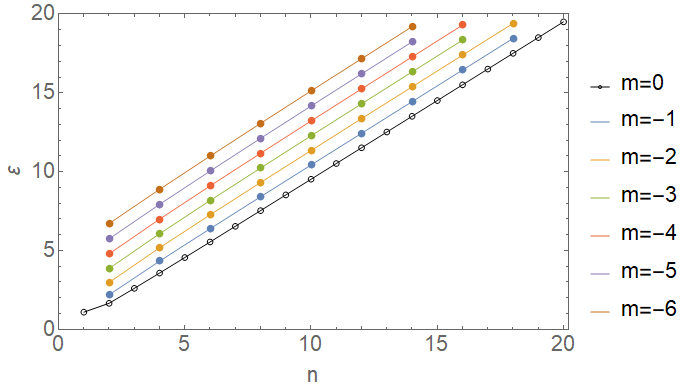}     
	\caption{$Q=1, m\leq0$}\label{EpMnQ1}
    \end{subfigure} 
    \begin{subfigure}{0.5\textwidth}
    \centering
	\includegraphics[scale=0.25]{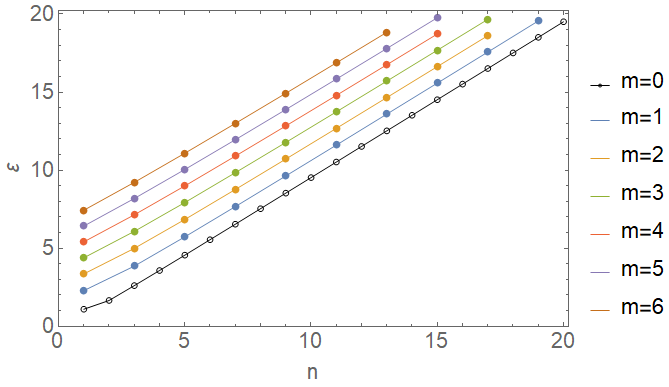}     
	\caption{$Q=1, m\geq0$}\label{EpMpQ1}
    \end{subfigure} 
    \\ 
	\\

    \begin{subfigure}{0.5\textwidth}
    \centering
    \includegraphics[scale=0.25]{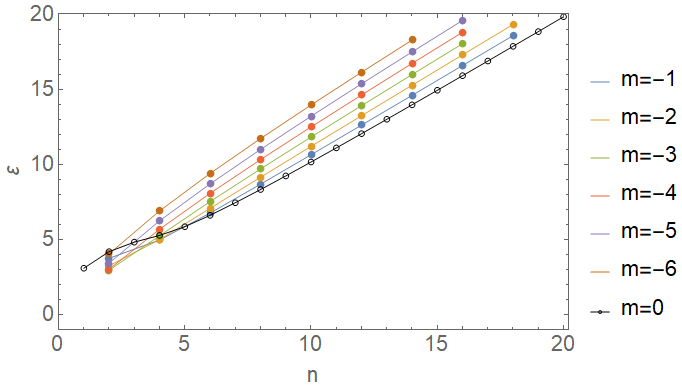} 
    \caption{$Q=5, m\leq0$}\label{EpMnQ5}
    \end{subfigure}
    \begin{subfigure}{0.5\textwidth}
    \centering
    \includegraphics[scale=0.25]{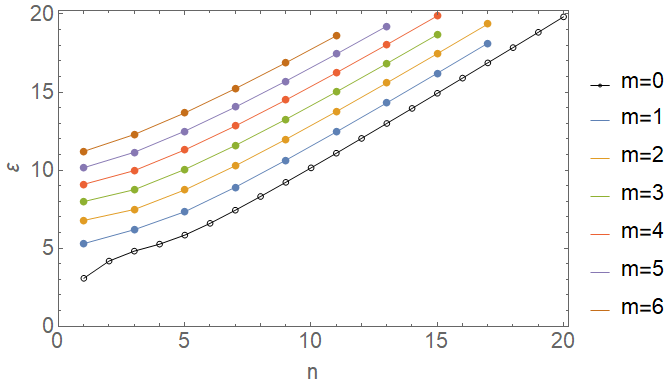} 
    \caption{$Q=5, m\geq0$}\label{EpMpQ5}
    \end{subfigure}
    \\
    \\
    
    \begin{subfigure}{0.5\textwidth}
    \centering
	\includegraphics[scale=0.25]{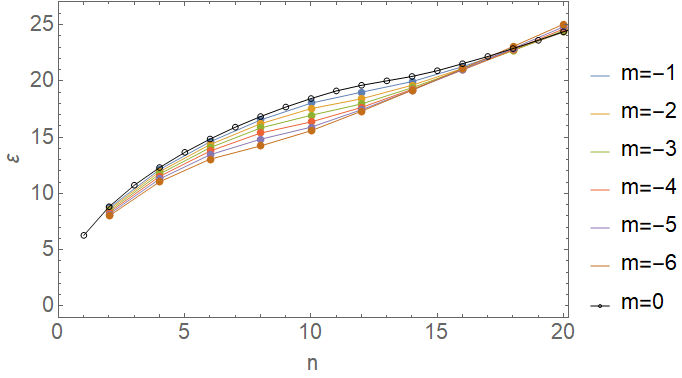} 
    \caption{$Q=20, m\leq0$}\label{EpMnQ20} 
    \end{subfigure} 
     \begin{subfigure}{0.5\textwidth}
    \centering
	\includegraphics[scale=0.25]{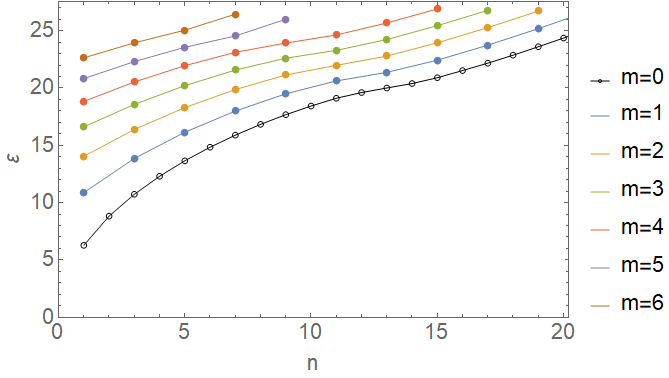} 
    \caption{$Q=20, m\geq0$}\label{EpMpQ20} 
    \end{subfigure} 
    \\
    \\

    \begin{subfigure}{0.5\textwidth}
    \centering
    \includegraphics[scale=0.25]{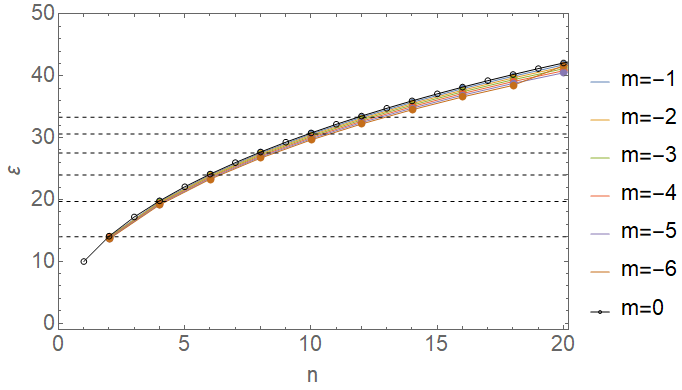} 
    \caption{$Q=50, m\leq0$}\label{EpMnQ50}
    \end{subfigure}
    \begin{subfigure}{0.5\textwidth}
    \centering
    \includegraphics[scale=0.25]{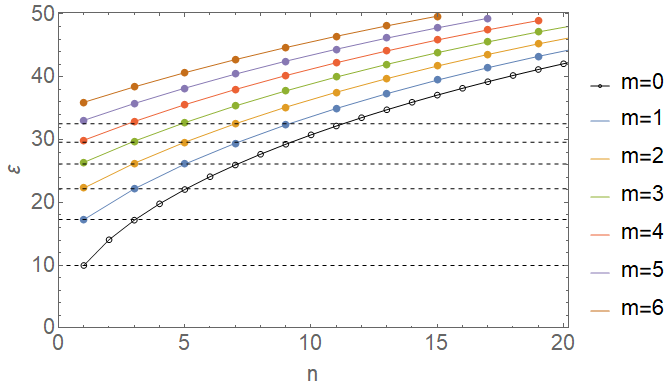} 
    \caption{$Q=50, m\geq0$}\label{EpMpQ50}
    \end{subfigure}
    
    \caption{Positive energy spectra for integer $m$.}\label{Ep}
\end{figure}

\begin{figure}[H]
    \begin{subfigure}{0.5\textwidth}
    \centering
	\includegraphics[scale=0.25]{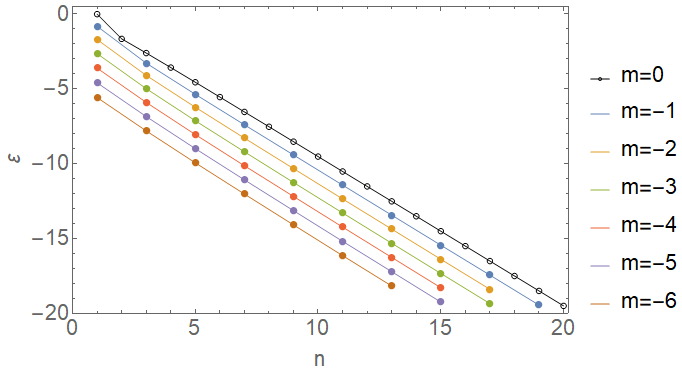}     
	\caption{$Q=1, m\leq0$}\label{EnMnQ1}
    \end{subfigure} 
     \begin{subfigure}{0.5\textwidth}
    \centering
	\includegraphics[scale=0.25]{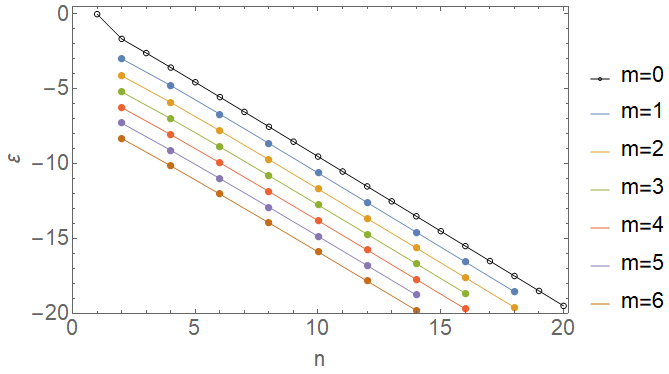}     
	\caption{$Q=1, m\geq0$}\label{EnMpQ1}
    \end{subfigure} 
    \\
    \\
    
    \begin{subfigure}{0.5\textwidth}
    \centering
    \includegraphics[scale=0.25]{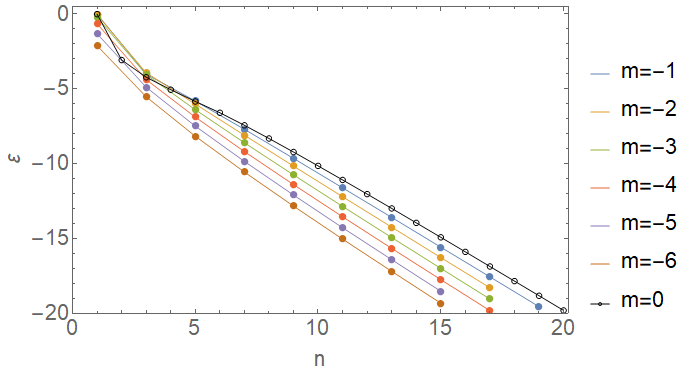} 
    \caption{$Q=5, m\leq0$}\label{EnMnQ5}
    \end{subfigure}
    \begin{subfigure}{0.5\textwidth}
    \centering
    \includegraphics[scale=0.25]{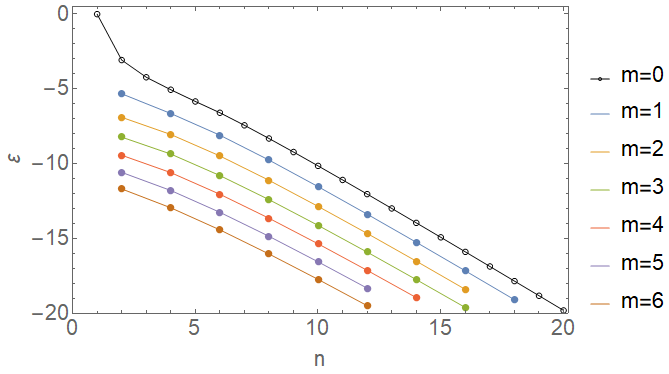} 
    \caption{$Q=5, m\geq0$}\label{EnMpQ5}
    \end{subfigure}
    \\
    \\
    
    \begin{subfigure}{0.5\textwidth}
    \centering
	\includegraphics[scale=0.25]{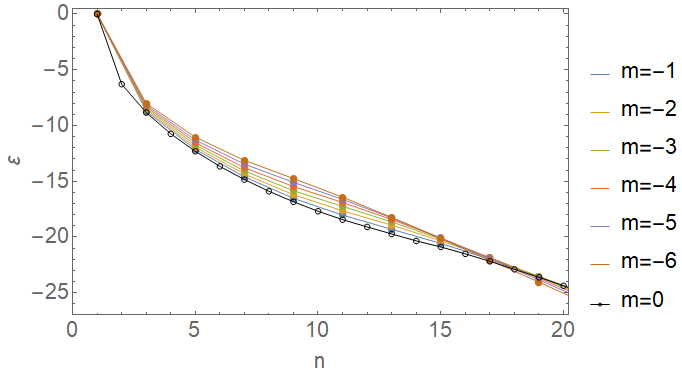} 
    \caption{$Q=20, m\leq0$}\label{EnMnQ20} 
    \end{subfigure} 
       \begin{subfigure}{0.5\textwidth}
    \centering
	\includegraphics[scale=0.25]{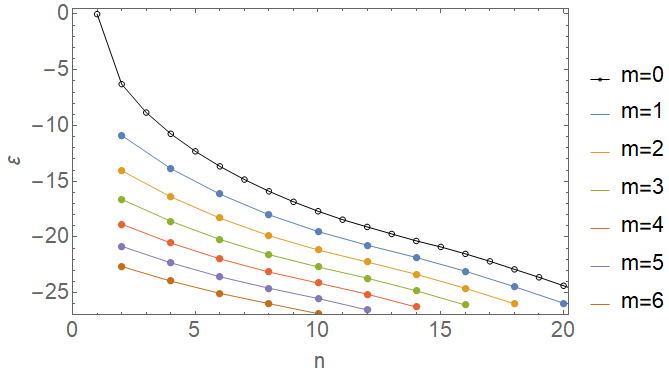} 
    \caption{$Q=20, m\geq0$}\label{EnMpQ20} 
    \end{subfigure} 
    \\
    \\
    
    \begin{subfigure}{0.5\textwidth}
    \centering
    \includegraphics[scale=0.25]{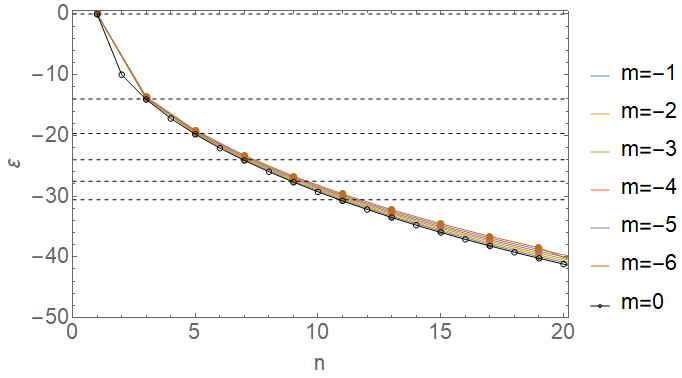} 
    \caption{$Q=50, m\leq0$}\label{EnMnQ50}
    \end{subfigure}
    \begin{subfigure}{0.5\textwidth}
    \centering
    \includegraphics[scale=0.25]{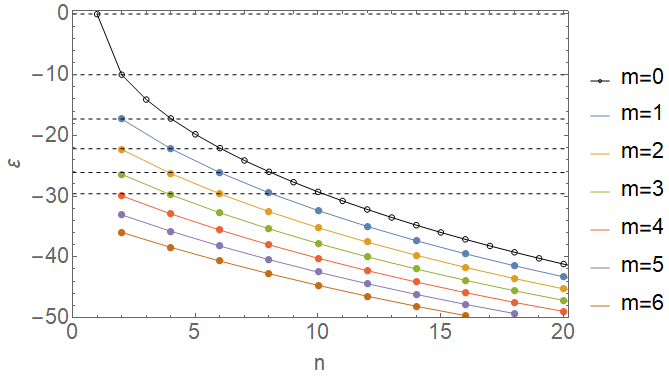} 
    \caption{$Q=50, m\geq0$}\label{EnMpQ50}
    \end{subfigure}
    \caption{Negative energy spectra for integer $m$.}\label{En}
\end{figure}

Finally, let's consider the case where $m\in \mathbb{Z}$, and compute the positive (Figure \ref{Ep}) and negative (Figure \ref{En}) spectra for various values of the momentum and dipole strength. Note that for fixed $Q$ the spectrum for $m=0$ takes on twice as many values within the same energy interval as the spectra for $m<0$ and $m>0$. The spectra for both $m<0$ and $m>0$ behave qualitatively as before, and our prior observations still hold. However, they relate to the $m=0$ spectra in a qualitatively new way. In the large $Q$ limit:
\begin{itemize}
\item The positive spectra are computed numerically and plotted in Figure \ref{Ep}. Notice that the $m<0$ energies merge into each other and into the $m=0$ energies with even $n$. The $m>0$ energies on the other hand are coincident with each other and with the $m=0$ energies with odd $n$. This is why we have chosen to plot the $m\neq0$ spectra only for every second\footnote{Note that this means that the quantum number $n$ which labels successive allowed energies only applies to the $m=0$ energies here.} $n$ value. 
\item Similarly, the negative spectra, plotted in Figure \ref{En}, show that $m<0$ states merge into each other and with $m=0$ states corresponding to odd $n$ while positive momentum states are coincident with each other and with $m=0$ states corresponding to even $n$.
\end{itemize}
The new feature of the overall spectrum in the large $Q$ limit as compared to the half-odd $m$ case is that the $m>0$ and $m<0$ levels no longer coincide, they now alternate resulting in twice as many Landau levels as before\footnote{When $Q\rightarrow\infty$ we expect there to be infinitely many Landau levels, so technically the statement should be that the \textit{density} of levels doubles.}. Finally, we also note that at small $n$, the spectrum again exhibits level crossing among the various $m\leq0$ states as $Q$ is increased.

\section{Conclusions}

The Hilbert space of a spinless quantum particle confined to a sphere enclosing a magnetic dipole is made up of the angular oblate spheroidal wavefunctions, labelled by integer quantum numbers $m$ and $l$. Unlike the Haldane sphere, Gauss' law renders the problem topologically trivial. Nevertheless, it exhibits a rich spectral structure, which is of particular interest in the limit $Q\gg 1$, in which it may be summarised as follows: 
\begin{itemize}
\item In the $|m|\ll Q$ regime, energy levels display an approximate, evenly spaced Landau level structure and states are localised about the poles. Additionally, states with pairwise adjacent $l-|m|$ values have degenerate energy.
\item In the $|m|\gg Q$ regime, energy levels display a quadratic dependence on $|m|$ and states are localised sharply about the equator. 
\item In the intermediate regime, the localisation of the wavefunctions drifts from the poles to the equator as $|m|$ increases. The spectrum has interesting structure only for $m<0$, where we find the system's lowest energy states, which satisfy $l=|m|\approx Q$ and are localised weakly about the equator.  
\end{itemize}
In this paper we have computed the corresponding Hilbert space for the spin-$\tfrac{1}{2}$ case, which may be summarised as follows: 
\begin{itemize}
    \item The two spinor component wavefunctions are related by reflection about the equator, and exhibit the same qualitative behaviour in the large $Q$ limit as the spinless wavefunctions.
    \item The same is true for the spectrum in the large $Q$ limit, with the exceptions that the Landau levels are no longer evenly spaced; the pairwise degeneracy in $l-|m|$ is now broken (with each degenerate branch taking opposite sign); and, we now see the existence of an approximately zero-energy Lowest landau level which persists past the $|m|\ll Q$ regime into the regime $|m|<Q$.
\end{itemize}
Let's now contextualize these results and speculate on some applications to condensed matter and gravitational physics.
\begin{itemize}
    \item A resurgence of interest in wormholes in the context of 2-dimensional gravity has led to some interesting developments in 4-dimensional magnetically charged black holes \eqref{magBH}, and their relation to traversable wormholes in four dimensions \cite{Maldacena:2020skw,Maldacena:2018gjk}. Key to stabilising such wormholes is the Landau degeneracy that accompanies the dynamics of massless charged fermions in the presence of a magnetic field on the sphere. The energy in each level receives contributions from orbital motion as well as a magnetic dipole contribution which, for a fermion in the lowest Landau level, exactly cancel as a consequence of the cancellation of the 2-dimensional gravitational anomaly. The result is a large set of states - corresponding to the large $q$ degeneracy of the lowest Landau level - with zero energy on the sphere giving rise to an equally large set of massless 2-dimensional chiral fermions in the $(r,t)$-directions. It is precisely the Casimir energy of these compact fermions that stabilizes the wormhole.\\
    
    \noindent
    Specifically, on the background \eqref{magBH} and in global coordinates, the representation $\gamma^{1} = i\sigma_{x}\otimes \mathbb{I}, \gamma^{2} = \sigma_{y}\otimes \mathbb{I}, \gamma^{3} = \sigma_{z}\otimes\sigma_{x}$ and $\gamma^{4} = \sigma_{z}\otimes\sigma_{y}$ for the $\gamma$-matrices, together with the ansatz $\chi_{\alpha\beta} = \psi_{\alpha}\otimes \eta_{\beta}$ for the 4-dimensional spinor in terms of the 2-dimensional spinors $\eta$ on the $S^{2}$ and $\psi$ in the remaining two directions, factorizes the 4-dimensional Dirac equation into a free massless Dirac equation for $\psi$ and a Dirac equation on the 2-sphere with magnetic field $(\slashed{\nabla} - i\slashed{A})\eta = 0$.
    The relevant geometry for the two mouths of the wormhole are a {\it pair} of oppositely charged magnetic black holes which produces a {\it dipole} vector potential of the kind that is the subject of this article. The authors of \cite{Maldacena:2018gjk} show that at distances comparable to the distance between the sources, the fermion wavefunctions localize on the field lines. This is sufficient for their purposes. However, when the distances is much larger than between the magnetic sources, our analysis above should be more appropriate. In particular, as we have shown, this would have implications for the lifting of the Landau degeneracy as well as the localization of the fermion wavefunctions, both of which, in turn, have consequences for the wormhole construction. 
    
    \item Modulo finite-size effects, graphene - a quasi-planar sheet of carbon atoms arranged in a hexagonal lattice and that exhibits some of the most remarkable known electronic and tensile properties - shares its electronic spectrum with the Dirac equation in (2+1)-dimensions. In particular, single electron dynamics in a graphene sample is captured by massless QED$_{2}$ on $\mathbb{R}^{2,1}$. This remains true for local (geometric) deformations of the graphene sheet. However, to form a spherical ball of graphene - a fullerine molecule like $\mathsf{C}_{60}$ - Euler's theorem requires the addition of pentagonal defects into the hexagonal graphene lattice; 12 pentagons in the case of $\mathsf{C}_{60}$. The effect of these defects can be encoded in the continuum field theory through the introduction of a fictitious magnetic monopole with fractional magnetic charge \cite{Gonzalez:1992qn}. The spectrum is obtained by solving the eigenvalue problem, $i\gamma^{\mu}\left(\partial_{\mu} + \Omega_{\mu} - iW_{\mu}\right)\Psi_{n} = E_{n}\Psi_{n}$, with spin connection $\Omega_{\mu} = \frac{1}{8}\omega^{\alpha\beta}_{\mu}[\gamma_{\alpha},\gamma_{\beta}]$ and $(W_{\theta},W_{\phi}) = (0,g\cot\theta)$ is the gauge connection for a Dirac monopole. Inserting the fullerene into a constant external magnetic field has the effect of modifying the Dirac equation to $i\gamma^{\mu}\left(\partial_{\mu} + \Omega_{\mu} - iW_{\mu} - iA_{\mu}\right)\Psi_{n} = E_{n}\Psi_{n}$, where, if the sphere is taken to have radius $R$, the only non-zero component of the gauge field is $A_{\phi} = \frac{1}{2}B_{0}R \sin\theta$. While the physics of this system is different from the dipole case - a spherical fullerene in a constant external field versus a fullerene enclosing a current loop in the latter - mathematically, for a fixed size sphere, they are identical. In this case also an exact solution can be found in term of oblate spheroidal functions \cite{Aoki}.\\
    
    \noindent
    The effect of this external constant magnetic field is two-fold; in the weak field limit ($Q_{\mathrm{const.}}\ll 1$) the spectrum exhibits orbital Zeeman splitting while the large $Q$ limit ($Q_{\mathrm{const.}}\gg 1$) is dominated by a Landau regime. At the crossover point between these two regimes, the spectrum exhibits a series of overlapping bands. Again, this is consistent with our findings in the dipole case. This is unsurprising since, for a fixed size of the sphere, both gauge potentials are functionally the same. In the constant-field case, since $Q_{\mathrm{const.}} = eB_{0}^{2}R^{2}/2$, the large $Q$ regime can be accessed by either turning up the strength of the magnetic field or increasing the size of the sphere (since the particle `sees' more of the magnetic field). In our dipole case however, since $Q_{\mathrm{dipole}} = e|\bm{\mu}|/R$, increasing the size of the sphere $\it decreases$ the effective magnetic charge. Consequently, increasing the cluster size from\footnote{For topological reasons, this is also accompanied by an increase in the strength of the monopole field. However, this is small relative to the parametrically dialled external field.} $\mathsf{C}_{60}$ to $\mathsf{C}_{240}$ in a fixed external $B_{0}$ increases $Q_{\mathrm{const}}$ and drives the system toward more defined Landau quantization, even as the continuum approximation improves. On the other hand, increasing the size of a spherical fullerene housing a current loop should have the opposite effect; the accompanied decrease in $Q_{\mathrm{dipole}}$ leads to enhanced Zeeman-like splitting in the spectrum (see, for example Figure \ref{HEpz}). The 0.7nm diameter of the $\mathsf{C}_{60}$ fullerene sets a natural scale for this problem and, given the recent leaps in nano-technology, it seems to us that fabrication of such a system of a fullerine enclosing a current loop is just around the proverbial corner. It would offer an exciting possibility to test some of these ideas.
\end{itemize}

\acknowledgments
J.M. was supported in part by the NRF of South Africa under grant CSUR 114599 and in part by a Simons Associateship at the ICTP, Trieste. R.P.S. is supported by a graduate fellowship from the National Institute for Theoretical Physics and by the Shuttleworth Postgraduate Scholarship Programme.


\end{document}